\documentclass[11pt]{article}
\usepackage{amsmath, amssymb, amsfonts,euscript,mathrsfs}
\usepackage{bm,bbm,latexsym,amsthm}
\usepackage{epsfig, verbatim,enumerate, wrapfig}
\usepackage[round]{natbib}
\usepackage{url}

\newcommand{\nc}{\newcommand}
\nc{\nn}{\nonumber}
\nc{\fns}{\footnotesize}

\nc{\revisionline}{\vspace{.1in} \today \vspace{.1in} \hrule\hrule\hrule\vspace{.1in}}
\nc{\newpp}{\vspace{.1in} \noindent}

\nc{\slideline}{\smallskip \hrule\hrule \smallskip}

\nc{\wh}{\widehat}
\nc{\wt}{\widetilde}

\def\boxit#1{\vbox{\hrule\hbox{\vrule\kern6pt
          \vbox{\kern6pt#1\kern6pt}\kern6pt\vrule}\hrule}}

\nc{\Ef}{ {\rm E}_{\infty} }
\nc{\Ex}{ {\rm E} }
\nc{\Ec}{ {\rm E}_1 }
\nc{\Pf}{ {\rm P}_{\infty} }
\nc{\Pc}{ {\rm P}_{1} }
\nc{\Prb}{ {\rm P} }
\nc{\sd}{\pm \hat{\sigma} }

\nc{\cond}{{\, \vert \,}}
\nc{\indep}{{\, \perp \! \! \! \perp  \,} }
\nc{\tsps}{^{ {\rm T} } }

\nc{\pu}{\pi_{\rm U}}
\nc{\pbi}{\pi_{\rm B}}
\nc{\pnb}{\pi_{\rm NB}}
\nc{\prp}{\propto}
\nc{\pr}{ {\rm pr} }

\nc{\texthalf}{\textstyle{1 \over 2}}
\nc{\textthird}{\textstyle{1 \over 3}}

\nc{\al}{\alpha}
\nc{\dl}{\delta}
\nc{\la}{\lambda}
\nc{\vep}{\varepsilon}
\nc{\eps}{\epsilon}

\nc{\snf}{\sum_{n=1}^{\infty}}
\nc{\skf}{\sum_{k=1}^{\infty}}
\nc{\sner}{\sum_{n=1}^{86}}
\nc{\sjn}{\sum_{j=1}^{n}}
\nc{\skn}{\sum_{k=1}^{n}}
\nc{\sumim}{\sum_{i=1}^m}
\nc{\sumjn}{\sum_{j=1}^n}
\nc{\sumlL}{\sum_{l=1}^{L}}
\nc{\sumL}{\sum_{l=1}^{L}}
\nc{\sumkK}{\sum_{k=1}^{K_i}}
\nc{\sumrR}{\sum_{r=1}^R}

\nc{\hivp}{\sum_{ {\rm HIV}^+ } }
\nc{\sumiN}{ \sum_{i=1}^N }
\nc{\summM}{ \sum_{m=1}^M }
\nc{\sumjM}{ \sum_{j=1}^M }
\nc{\lsq}{\left[}
\nc{\rsq}{\right]}
\nc{\lbc}{\left \{ }
\nc{\rbc}{\right \} }
\nc{\lp}{\left(}
\nc{\rp}{\right)}

\nc{\imp}{\Rightarrow}
\nc{\lbf}{\lim_{b \rightarrow \infty}}
\nc{\limNinf}{\lim_{N \rightarrow \infty}}
\nc{\limminf}{\lim_{m \rightarrow \infty}}
\nc{\limninf}{\lim_{n \rightarrow \infty}}
\nc{\convd}{\stackrel{D}{\longrightarrow}}
\nc{\convp}{\stackrel{P}{\longrightarrow}}
\nc{\eqd}{\stackrel{{\EuScript D}}{=}}

\nc{\subi}{_{\text{I}}}
\nc{\subs}{_{\text{S}}}
\nc{\subni}{_{\text{NI}}}

\nc{\trans}{^{\text T}}
\nc{\ol}{\overline}

\DeclareMathOperator{\var}{var}

\nc{\rl}{ {\rm {\bf R} } }
\nc{\zah}{ {\rm {\bf Z} } }

\nc{\lkn}{\Lambda^n_k}
\nc{\stp}{ {\cal C}_b }
\nc{\istp}{ {\cal I}_A }
\nc{\snb}{S_{N_b}}
\nc{\stb}{S_{T_b}}
\nc{\ixlog}{I_{ \{ 0 \leq x \leq \log \al \} } }
\nc{\iulog}{I_{ \{ 0 \leq u  \leq \log \al \} } }
\nc{\rgn}{ \Upsilon_n }
\nc{\dpl}{\partial}
\nc{\half}{ {\textstyle \frac{1}{2}} }
\nc{\real}{\mathbb{R}}
\nc{\bbC}{\mathbb{C}}
\nc{\bbR}{\mathbb{R}}
\nc{\bbone}{\mathbb{1}}
\nc{\bbP}{\mathbb{P}}

\def\boxit#1{\vbox{\hrule\hbox{\vrule\kern6pt\vbox{\kern6pt#1\kern6pt}\kern6pt\vrule}\hrule}}

\nc{\calb}{ {\cal B} }
\nc{\calc}{ {\cal C} }
\nc{\bcalc}{ \mbox{\boldmath{${\cal C}$}}}
\nc{\cald}{ {\cal D} }
\nc{\cale}{ {\cal E} }
\nc{\cali}{ {\cal I} }
\nc{\call}{ {\cal L} }
\nc{\calm}{ {\cal M} }
\nc{\caln}{ {\cal N} }
\nc{\cals}{ {\cal S} }
\nc{\calo}{ {\cal O} }
\nc{\bcalo}{ \mbox{\boldmath{${\cal O}$}}}
\nc{\calt}{ {\cal T} }
\nc{\calv}{ {\cal V} }
\nc{\bcalu}{ \mbox{\boldmath{${\cal U}$}}}
\nc{\calu}{ {\cal U} }
\nc{\calw}{ {\cal W} }
\nc{\calx}{ {\cal X} }

\nc{\sca}{ {\EuScript A} }
\nc{\scb}{ {\EuScript B} }
\nc{\scc}{ {\EuScript C} }
\nc{\scd}{ {\EuScript D} }
\nc{\sce}{ {\EuScript E} }
\nc{\scf}{ {\EuScript F} }
\nc{\scF}{ {\EuScript f} }
\nc{\scg}{ {\EuScript G} }
\nc{\sch}{ {\EuScript H} }
\nc{\sci}{ {\EuScript I} }
\nc{\scj}{ {\EuScript J} }
\nc{\sck}{ {\EuScript K} }
\nc{\scl}{ {\EuScript L} }
\nc{\sclic}{ \scl_i^{\rm c} }
\nc{\scm}{ {\EuScript M} }
\nc{\scn}{ {\EuScript N} }
\nc{\sco}{ {\EuScript O} }
\nc{\scp}{ {\EuScript P} }
\nc{\scq}{ {\EuScript Q} }
\nc{\scr}{ {\EuScript R} }
\nc{\scs}{ {\EuScript S} }
\nc{\sct}{ {\EuScript T} }
\nc{\scu}{ {\EuScript U} }
\nc{\scv}{ {\EuScript V} }
\nc{\scw}{ {\EuScript W} }
\nc{\scx}{ {\EuScript X} }
\nc{\scy}{ {\EuScript Y} }
\nc{\scz}{ {\EuScript Z} }
\nc{\scxo}{ {\EuScript X}_{\rm obs} }
\nc{\Xobs}{ \pmb{\scx}_{\rm obs} }
\nc{\Xcom}{ \pmb{\scx} }
\nc{\Xmis}{ \pmb{\scx}_{\rm mis} }

\nc{\bsci}{ \mbox{\boldmath{$\sci$}}}
\nc{\bscj}{ \mbox{\boldmath{$\scj$}}}

\nc{\sumlic}{\sum_{l \in sclic}}

\nc{\scyo}{ {\EuScript Y}_{\rm obs} }

\nc{\bga}{\begin{array}{c}}
\nc{\ena}{\end{array}}

\nc{\mhat}{ {\hat{p}}_M }
\nc{\fhat}{ {\hat{p}}_F }
\nc{\ph} { \hat{p} }

\nc{\redDelta}{\textcolor{red}{\Delta}}
\nc{\redbeta}{\textcolor{red}{\beta}}

\nc{\red}{\textcolor{red}}
\nc{\blue}{\textcolor{blue}}

\nc{\ta}{ {\tilde{a}} }
\nc{\tc}{ {\tilde{c}} }

\nc{\balpha}{\pmb{\alpha}}
\nc{\bal}{\pmb{\alpha}}
\nc{\bone}{\pmb{1}}
\nc{\bbet}{\pmb{\beta}}
\nc{\bbeta}{\pmb{\beta}}
\nc{\bDel}{\pmb{\Delta}}
\nc{\bDelta}{\pmb{\Delta}}
\nc{\bdel}{\pmb{\delta}}
\nc{\bdelta}{\pmb{\delta}}
\nc{\bet}{\pmb{\eta}}
\nc{\beps}{\pmb{\epsilon}}
\nc{\bvep}{\pmb{\vep}}
\nc{\bgam}{\pmb{\gamma}}
\nc{\bgamma}{\pmb{\gamma}}
\nc{\bGamma}{\pmb{\Gamma}}
\nc{\boldeta}{\pmb{\eta}}
\nc{\bLam}{\pmb{\Lambda}}
\nc{\bLambda}{\pmb{\Lambda}}
\nc{\blambda}{\pmb{\lambda}}
\nc{\bmu}{ \pmb{\mu}}
\nc{\boldnu}{ \pmb{\nu}}
\nc{\bOm}{ \pmb{\Omega}}
\nc{\bOmega}{ \pmb{\Omega}}
\nc{\bom}{ \pmb{\omega}}
\nc{\bomega}{ \pmb{\omega}}
\nc{\bpi}{ \pmb{\pi}}
\nc{\bPi}{ \pmb{\Pi}}
\nc{\bpsi}{ \pmb{\psi}}
\nc{\bPsi}{ \pmb{\Psi}}
\nc{\bphi}{ \pmb{\phi}}
\nc{\bPhi}{ \pmb{\Phi}}
\nc{\bxi}{ \pmb{\xi}}
\nc{\bXi}{ \pmb{\Xi}}
\nc{\bSig}{\pmb{\Sigma}}
\nc{\bSigma}{\pmb{\Sigma}}
\nc{\bsig}{\pmb{\sigma}}
\nc{\bsigma}{\pmb{\sigma}}
\nc{\btau}{\pmb{\tau}}
\nc{\bThe}{\pmb{\Theta}}
\nc{\bTheta}{\pmb{\Theta}}
\nc{\bthe}{\pmb{\theta}}
\nc{\btheta}{\pmb{\theta}}
\nc{\bzeta}{\pmb{\zeta}}
\nc{\bIm}{\pmb{\Im}}

\nc{\ba}{ \pmb{ a }}
\nc{\bA}{ \pmb{ A }}
\nc{\bB}{ \pmb{ B }}
\nc{\bb}{ \pmb{ b }}
\nc{\bc}{ \pmb{ c }}
\nc{\bC}{ \pmb{ C }}
\nc{\bD}{ \pmb{ D }}
\nc{\bd}{ \pmb{ d }}
\nc{\be}{ \pmb{ e }}
\nc{\bF}{ \pmb{ F }}
\nc{\boldf}{ \pmb{ f }}
\nc{\bg}{ \pmb{ g }}
\nc{\bG}{ \pmb{ G }}
\nc{\bh}{ \pmb{ h }}
\nc{\bH}{ \pmb{ H }}
\nc{\bI}{ \pmb{ I }}
\nc{\bJ}{ \pmb{ J }}
\nc{\bK}{ \pmb{ K }}
\nc{\bk}{ \pmb{ k }}
\nc{\bL}{ \pmb{ L }}
\nc{\bM}{ \pmb{ M }}
\nc{\bn}{ \pmb{ n }}
\nc{\bO}{ \pmb{ O }}
\nc{\bP}{ \pmb{ P }}
\nc{\bp}{ \pmb{ p }}
\nc{\bQ}{ \pmb{ Q }}
\nc{\br}{ \pmb{ r }}
\nc{\bR}{ \pmb{ R }}
\nc{\bolds}{ \pmb{ s }}
\nc{\bS}{ \pmb{ S }}
\nc{\bT}{ \pmb{ T }}
\nc{\bt}{ \pmb{ t }}
\nc{\bu}{ \pmb{ u }}
\nc{\bU}{ \pmb{ U }}
\nc{\bv}{ \pmb{ v }}
\nc{\bV}{ \pmb{ V }}
\nc{\bW}{ \pmb{ W }}
\nc{\bw}{ \pmb{ w }}
\nc{\bx}{ \pmb{ x }}
\nc{\bX}{ \pmb{ X }}
\nc{\by}{ \pmb{ y }}
\nc{\bY}{ \pmb{ Y }}
\nc{\bz}{ \pmb{ z }}
\nc{\bZ}{ \pmb{ Z }}

\nc{\YR}{[\bY,R]}
\nc{\YgivenR}{[\bY \mid R]}
\nc{\RgivenY}{[R \mid \bY]}
\nc{\Y}{[\bY]}
\nc{\redmuzero}{\textcolor{red}{\mu_0}}

\nc{\dio}{d_i^o}
\nc{\timi}{t_{i,m_i}}
\nc{\bbetahat}{\hat{\bbet}}
\nc{\mui}{\bmu_{\rm I}}
\nc{\mue}{\bmu^{\rm E}}
\nc{\mup}{\bmu^{\rm P}}
\nc{\muihat}{\hat{\bmu}_{\rm I}}
\nc{\muehat}{\hat{\bmu}^{\rm E}}
\nc{\muphat}{\hat{\bmu}^{\rm P}}
\nc{\bdelhat}{\hat{\bdel}}
\nc{\bmuhat}{\hat{\bmu}}

\nc{\iid}{\stackrel{\rm iid}{\sim}}
\nc{\law}{\stackrel{\scl}{=}}

\nc{\phiij}{ \phi_{ij}( \Delta_0) }
\nc{\phiiprmj}{ \phi_{i'j}( \Delta_0) }
\nc{\phiijprm}{ \phi_{ij'}( \Delta_0) }
\nc{\phixy}{ \phi( X_i(S_{ik}), Y_j(T_{jl}) ) }
\nc{\phixydo}{ \phi( X_i(S_{ik}), Y_j(T_{jl})-\Delta_0 ) }
\nc{\phixyd}{ \phi( X_i(S_{ik}), Y_j(T_{jl})-\Delta) }
\nc{\phixydstar}{ \phi^*( X_i(S_{ik}), Y_j(T_{jl})-\Delta) }
\nc{\phixystdttil}{ \tilde{\phi}( X_i(s), Y_j(t)-\Delta, \theta) }
\nc{\phixydttil}{ \tilde{\phi}( X_i(S_{ik}), Y_j(T_{jl})-\Delta, \theta) }
\nc{\Nmn}{{\sqrt{N} \over mn}}

\nc{\Xis}{X_i(s)}
\nc{\Yjt}{Y_j(t)}

\nc{\bthehat}{\hat{\bthe}}

\nc{\alphahat}{\wh{\alpha}}
\nc{\betahat}{\wh{\beta}}
\nc{\deltahat}{\wh{\delta}}
\nc{\gammahat}{\wh{\gamma}}
\nc{\thetahat}{\wh{\theta}}
\nc{\phihat}{\wh{\phi}}
\nc{\psihat}{\wh{\psi}}
\nc{\pihat}{\wh{\pi}}
\nc{\muhat}{\wh{\mu}}
\nc{\etahat}{\wh{\eta}}

\nc{\Yhat}{\wh{Y}}
\nc{\Fhat}{\wh{F}}

\nc{\Ghat}{\what{G}}

\nc{\Ehat}{\wh{E}}
\nc{\varhat}{\wh{\var}}

\nc{\Ritil}{\tilde{R}_i}

\nc{\Ybar}{\overline{Y}}
\nc{\Rbar}{\overline{R}}
\nc{\Nbar}{\overline{N}}
\nc{\intzeroinf}{\int_0^\infty}

\nc{\FhatS}{\hat{F}(S_{ik})}
\nc{\GhatT}{\hat{G}(T_{jl})}

\nc{\Fhatik}{\hat{F}_{ik}}
\nc{\Ghatjl}{\hat{G}_{jl}}
\nc{\Fik}{F_{ik}}
\nc{\Gjl}{G_{jl}}
\nc{\phiijkl}{\phi_{ik,jl}(\Delta)}
\nc{\phiijkltil}{\tilde{\phi}_{ik,jl}(\Delta_0,\theta_0)}
\nc{\ord}{N^{-3/2}}           % --- effective order of summation
\nc{\sumijkl}{\sum_{ijkl}}

\nc{\Citil}{\tilde{C}_i}
\nc{\Crtil}{\tilde{C}_r}
\nc{\Djtil}{\tilde{D}_j}
\nc{\Ditil}{\tilde{D}_i}

\nc{\Cithe}{\tilde{C}^{\theta}_i}
\nc{\Djthe}{\tilde{D}^{\theta}_j}

\nc{\Sikthe}{S_{ik}^{\theta}}
\nc{\Tjlthe}{T_{jl}^{\theta}}

\nc{\Zi}{ \bZ_{-i}}
\nc{\zic}{ \lbc z(\bs_j) \: : \: i \neq j \rbc }
\nc{\zkap}{ \bz_{\kappa} }
\nc{\sumi}{ \sum_i }
\nc{\sumj}{ \sum_j }
\nc{\sumij}{ \sum_{i < j} }
\nc{\sumiandj}{ \sum_{i, j} }
\nc{\zsi}{ z(\bs_i) }
\nc{\Zsi}{ Z(\bs_i) }
\nc{\zsj}{ z(\bs_j) }
\nc{\zsn}{ z(\bs_n) }
\nc{\zsone}{ z(\bs_1) }
\nc{\pZ}{ \Pr \lbc \bZ \rbc }
\nc{\qz}{ Q( \bz ) }
\nc{\qZ}{ Q( \bZ ) }

\nc{\thetaYD}{\theta_{Y\mid D}}
\nc{\thetaD}{\theta_D}
\nc{\psiDY}{\psi_{D\mid Y}}
\nc{\psiY}{\psi_Y}

\nc{\tn}{\Theta^{\nu}}
\nc{\Etn}{E_{\theta^{\nu}}}
\nc{\tnone}{\Theta^{\nu+1}}
\nc{\Lm}{L_{\text{m}}}
\nc{\Lo}{L_{\text{o}}}
\nc{\Ym}{Y_{\text{m}}}
\nc{\Yo}{Y_{\text{o}}}
\nc{\ym}{y_{\text{m}}}
\nc{\yo}{y_{\text{o}}}

\nc{\vijb}{v_{ij} - \bX_{i(j)}  \bbet}
\nc{\vikb}{v_{ik} - \bX_{i(k)}  \bbet}
\nc{\vilb}{v_{il} - \bX_{i(l)}  \bbet}
\nc{\betart}{ \bbet^{(r)}_{t_i} }
\nc{\betarj}{ \bbet^{(r)}_j }
\nc{\yij}{y_{ij}}
\nc{\Xmisi}{ {\bX_{ i{\rm (mis)} }} }
\nc{\Xobsi}{ {\bX_{ i{\rm (obs)} }} }
\nc{\Zobsi}{ {\bZ_{ i{\rm (obs)} }} }
\nc{\bSigobs}{ \bSig_{  {\rm obs} } }
\nc{\bSigmis}{ \bSig_{  {\rm mis} } }
\nc{\bSigmo}{ \bSig_{  {\rm mis,obs} } }
\nc{\bSigom}{ \bSig_{  {\rm obs,mis} } }
\nc{\Xil}{{\bX}_{il}}
\nc{\Zil}{{\bZ}_{il} }
\nc{\omilr}{\omega_{il}^{(r)}}
\nc{\delio}{\bdel_i^{{\rm obs}} }

\nc{\obs}{{\text{obs}}}
\nc{\mis}{{\text{mis}}}
\nc{\rep}{{\text{rep}}}

\nc{\supone}{^{(1)}}
\nc{\supzero}{^{(0)}}
\nc{\supb}{^{(b)}}
\nc{\supr}{^{(r)}}
\nc{\supk}{^{(k)}}
\nc{\sups}{^{(s)}}
\nc{\supell}{^{(\ell)}}
\nc{\supgeqtwo}{^{(\geq 2)}}
\nc{\supthree}{^{(3)}}
\nc{\suptwo}{^{(2)}}
\nc{\subE}{_{\text{E}}}
\nc{\subO}{_{\text{O}}}
\nc{\subENI}{_{\text{E:NI}}}
\nc{\subEI}{_{\text{E:I}}}
\nc{\subNI}{_{\text{NI}}}
\nc{\subI}{_{\text{I}}}

\nc{\yio}{ {y_i^{\rm o} }}
\nc{\Yio}{ {Y_i^{\rm o }} }
\nc{\Yim}{ {Y_i^{\rm m} }}
\nc{\yim}{ {y_i^{\rm m} }}
\nc{\Yc}{Y^{\rm c}}
\nc{\Yic}{Y_i^{\rm c}}
\nc{\yc}{y^{\rm c}}
\nc{\yic}{y_i^{\rm c}}

\nc{\bYobsi}{\bY_{i,\text{obs}}}
\nc{\bYmisi}{\bY_{i,\text{mis}}}
\nc{\byobsi}{\by_{i,\text{obs}}}
\nc{\bymisi}{\by_{i,\text{mis}}}
\nc{\bymis}{\by_{\text{mis}}}
\nc{\byobs}{\by_{\text{obs}}}
\nc{\bYmis}{\bY_{\text{mis}}}
\nc{\bYobs}{\bY_{\text{obs}}}

\nc{\Ymis}{Y_{\text{mis}}}
\nc{\Yobs}{Y_{\text{obs}}}

\nc{\ymis}{y_{\text{mis}}}
\nc{\yobs}{y_{\text{obs}}}

\nc{\fyoym}{f_{Y_{\text{obs}},Y_{\text{\mis}}}}
\nc{\fyo}{f_{Y_{\text{obs}}}}
\nc{\fyoymr}{f_{Y_{\text{obs}},Y_{\text{\mis}},R}}
\nc{\fyor}{f_{Y_{\text{obs}},R} }

\nc{\yi}{y_i}
\nc{\Yi}{Y_i}

\nc{\fyic}{f ( \yic ; \; \psiY )}
\nc{\fyi}{f ( y_i ; \; \psiY ) }
\nc{\fdigivenyic}{f ( d_i  \mid  \yic ; \; \psiDY )}
\nc{\fditilgivenyic}{f ( \tilde{d}_i  \mid  \yic ; \; \psiDY )}
\nc{\fditilgivenyi}{f ( \tilde{d}_i  \mid  \yi ; \; \psiDY )}
\nc{\Fditilgivenyic}{F ( \tilde{d}_i  \mid  \yic ; \; \psiDY )}
\nc{\Fditilgivenyi}{F ( \tilde{d}_i  \mid  \yi ; \; \psiDY )}
\nc{\fdigivenyi}{f (d_i \mid y_i ; \;  \psiDY  )}
\nc{\fyicdi}{f \left( \yic, d_i \right)}
\nc{\fyidi}{f \left( \yi, d_i \right)}

\nc{\fymidr}{f_{Y \mid R}}
\nc{\fyr}{f_{Y,R}}
\nc{\frmidy}{f_{R \mid Y}}
\nc{\fy}{f_Y}
\nc{\fr}{f_R}

\nc{\fyicgivendi}{f (\yic \mid d_i; \; \thetaYD )}
\nc{\fyigivendi}{f (\yi \mid d_i; \; \thetaYD )}
\nc{\fyicgivens}{f (\yic \mid s; \; \thetaYD )}
\nc{\fyigivens}{f (\yi \mid s; \; \thetaYD )}
\nc{\fdi}{f ( d_i; \; \thetaD )}

\nc{\fyicX}{f ( \yic \mid X_i; \; \psiY )}
\nc{\fyiX}{f ( y_i \mid X_i; \; \psiY ) }
\nc{\fdigivenyicX}{f ( d_i  \mid  \yic, X_i ; \; \psiDY )}
\nc{\fdigivenyiX}{f (d_i \mid y_i, X_i ; \;  \psiDY  )}
\nc{\fyicdiX}{f \left( \yic, d_i \mid X_i \right)}

\nc{\fyicgivendiX}{f (\yic \mid d_i, X_i; \; \thetaYD )}
\nc{\fyigivendiX}{f (y_i \mid d_i, X_i; \; \thetaYD )}
\nc{\fdiX}{f ( d_i \mid X_i; \; \thetaD )}

\nc{\Yistar}{\bY_i^*}

\nc{\Dio}{D_i^{\rm obs}}
\nc{\bdelio}{\bdel_{ i \, {\rm (obs)}} }

\nc{\fygivend}{f_{Y \mid \delta}}
\nc{\fyd}{f_{Y, \delta}}
\nc{\fd}{f_\delta}
\nc{\FD}{F_D}
\nc{\fygivenbd}{f_{Y\mid b, \delta}}

\nc{\bthetahat}{\hat{\bthe}}
\nc{\thetatilde}{\tilde{\bthe}}
\nc{\scoretheta}{\bS(\bthe; \, \scc)}
\nc{\hesstheta}{\bH(\bthe; \, \scc)}
\nc{\infotheta}{\sci(\bthe; \, \scc)}
\nc{\sitheta}{\bs_i(\bthe; \, \scc_i)}
\nc{\sithetahat}{\bs_i(\thetahat; \, \scc_i)}

\nc{\loglikobs}{\ell_{{\rm o}}(\bthe; \, \sco)}
\nc{\scoreobs}{\bS_{{\rm o}}(\bthe; \, \sco)}
\nc{\hessobs}{\bH_{{\rm o}}(\bthe; \, \sco)}
\nc{\infoobs}{\scj_{{\rm o}}(\bthe; \, \sco)}

\nc{\Cil}{\scc_{il}}
\nc{\olog}{\lambda^*(\bthe, \Xobs)}
\nc{\LthetaC}{\scl(\bthe; \, \scc)}
\nc{\LthetaCi}{\scl_i(\bthe; \, \scc_i)}
\nc{\LthetaCil}{\scl_i (\bthe; \, \scc_{il}) }
\nc{\lthetaC}{\ell(\bthe; \, \scc)}
\nc{\lthetaCi}{\ell_i(\bthe; \, \scc_i)}
\nc{\lthetaCil}{\ell_i (\bthe; \, \scc_{il}) }
\nc{\Qtheta}{\scq \left( \bthe \, \left| \,  \bthe^{(r)} \right. \right)}
\nc{\thetar}{\bthe^{(r)}}
\nc{\thetas}{\bthe^{(s)}}
\nc{\alphas}{\bal^{(s)}}
\nc{\psis}{\psi^{(s)}}
\nc{\alphasplusone}{\bal^{(s+1)}}
\nc{\psisplusone}{\bpsi^{(s+1)}}
\nc{\alphapsis}{\left( \alphas, \psis \right)}

\nc{\thetarplusone}{\bthe^{(r+1)}}
\nc{\ologi}{\lambda^*_i(\bthe, \Xobs)}
\nc{\llogi}{\lambda_i \left( \bthe, \tilde{\Xcom}_{il} \right) }

\nc{\scxil}{\tilde{\Xcom}_{il}}
\nc{\siginv}{\bSig_i^{-1}}

\nc{\fofym}{ f \left( \by_i \mid \bbet_m, \bSig \right) }
\nc{\mphim}{ \phi_M \lsq \bSig^{-1/2}(\by_i - \bX_i \bbet_m) \rsq }
\nc{\mphit}{ \phi_M \lsq \bSig^{-1/2}(\by_i - \bX_i \bbet_{t_i}) \rsq }
\nc{\mphij}{ \phi_M \lsq \bSig^{-1/2}(\by_i - \bX_i \bbet_j) \rsq }
\nc{\mphik}{ \phi_M \lsq \bSig^{-1/2}(\by_i - \bX_i \bbet_k) \rsq }
\nc{\expkerm}{ \exp  \lbc -\half \bu_i(\bbet_m)' \bSig^{-1} \bu_i(\bbet_m)
  \rbc }
\nc{\expkerk}{ \exp  \lbc -\half \bu_i(\bbet_k)' \bSig^{-1} \bu_i(\bbet_k)
  \rbc }
\nc{\expkerj}{ \exp  \lbc -\half \bu_i(\bbet_j)' \bSig^{-1} \bu_i(\bbet_j)
  \rbc }
\nc{\normscorem}{\left( \bX_i' \bSig^{-1} \bX_i \bbet_m - \bX_i' \bSig^{-1}
  \by_i \right) }
\nc{\normscorej}{\left( \bX_i' \bSig^{-1} \bX_i \bbet_j - \bX_i' \bSig^{-1}
  \by_i \right) }
\nc{\piti}{ \pi \left( t_i, \bal, \bZ_i\bgam \right) }
\nc{\omij}{ \om_{ij} \left( t_i, \bal, \bZ_i\bgam \right) }
\nc{\phibetak}{ \phi_M(\bbet_k) }
\nc{\phibetaj}{ \phi_M(\bbet_j) }
\nc{\dphidbetak}{ \left. \dpl \phibetak \right/ \dpl \bbet_k }
\nc{\dphidbetakf}{ \frac{ \dpl \phibetak }{ \dpl \bbet_k } }
\nc{\uik}{\bu_i \left( \bbet_k  \right)}
\nc{\mset}{ \{ 0, 1, \ldots, M \} }
\nc{\betasigma}{ \left( \lbc \bbet^{(r)}_t \rbc, \bSig^{(r)} \right) }
\nc{\Thetar}{ \bThe^{(r)} }

\nc{\shatkm}{\hat{S}_{\rm KM}}

\nc{\ds}{\displaystyle}

\nc{\baln}{\begin{align*}}
\nc{\ealn}{\end{align*}}

\nc{\balna}{\begin{align}}
\nc{\ealna}{\end{align}}

\nc{\beq}{\begin{eqnarray*}}
\nc{\eeq}{\end{eqnarray*}}

\nc{\beqna}{\begin{eqnarray}}
\nc{\eeqna}{\end{eqnarray}}

\nc{\bct}{\begin{center}}
\nc{\ect}{\end{center}}

\nc{\bds}{\begin{description}}
\nc{\eds}{\end{description}}

\nc{\bit}{\begin{itemize}}
\nc{\eit}{\end{itemize}}

\nc{\bnu}{\begin{enumerate}}
\nc{\enu}{\end{enumerate}}

\nc{\bgt}{\begin{table}}
\nc{\bgtb}{\begin{center} \begin{tabular}}
\nc{\entb}{\end{tabular} \end{center} }
\nc{\ent}{\end{table}}

\nc{\ts}{\textstyle}

\nc{\bgs}{\begin{slide}}
\nc{\ens}{\end{slide}}

\nc{\hitem}{\vspace{.5in} \item}
\nc{\qitem}{\vspace{.25in} \item}
\nc{\titem}{\vspace{.75in} \item}
\nc{\fitem}{\vspace{1in} \item}
\nc{\eitem}{\vspace{.125in} \item}

\usepackage{mdwlist}

\usepackage[top=1in, bottom=1in, left=1.5in, right=1in]{geometry}

\usepackage{natbib}
\usepackage{graphicx}
\graphicspath{%
    {converted_graphics/}% inserted by PCTeX
    {/}% inserted by PCTeX
}
\usepackage{setspace}
\usepackage{titlesec}
\titleformat{\chapter}[display]
  {\normalfont\filcenter}{\MakeUppercase{{\chaptertitlename}}~\thechapter}{18pt}{}

\begin{document}
%%%%%%%%%%%

\noindent
\begin{center}
{  {\bf{Reduced Bias for respondent driven sampling: accounting for non-uniform edge sampling probabilities in people who inject drugs in Mauritius}}}        

\end{center}      

\doublespacing
\begin{abstract}
People who inject drugs are an important population to study in order to reduce transmission of blood-borne illnesses including HIV and Hepatitis.  In this paper we estimate the HIV and Hepatitis C prevalence among people who inject drugs, as well as the proportion of people who inject drugs who are female  in Mauritius. Respondent driven sampling (RDS), a widely adopted link-tracing sampling design used to collect samples from hard-to-reach human populations, was used to collect this sample.  The random walk approximation underlying many common RDS estimators assumes that each social relation (edge) in the underlying social network has an equal probability of being traced in the collection of the sample.  This assumption does not hold in practice.  We show that certain RDS estimators are sensitive to the violation of this assumption.  In order to address this limitation in current methodology, and the impact it may have on prevalence estimates, we present a new method for improving RDS prevalence estimators using estimated edge inclusion probabilities, and apply this to data from Mauritius. \\
 
\end{abstract}
KEY WORDS: Respondent Driven Sampling, Link-Tracing, Network Sampling, Edge Inclusion, HIV, AIDS, Hepatitis\\ %Homophily\\

\noindent Authors: Miles Q. Ott, Krista J. Gile, Matthew T. Harrison, Lisa G. Johnston, Joseph W. Hogan\\

\noindent Corresponding Author: Miles Q. Ott, mott@smith.edu, Statistical and Data Sciences Program, Smith College, 7 College Lane, Northampton MA 01063 

\newpage
%\tableofcontents 

\section{Introduction}\label{intro}

\subsection{Injection Drug Use in Mauritius}

Mauritius is estimated to have one of the highest per capita percentages of people who inject drugs (PWID) of all African countries \citep{Johnston2013sextransm,Mauritius2014}.  This high rate of injection drug use has seriously impacted public health, as it is the primary mode of HIV transmission within Mauritius, and accounts for 44\% of all HIV transmissions in the country \citep{Mauritius2015}. In order to measure HIV and other infections' prevalence and associated risk factors in Mauritius, a sample of 500 PWID was collected using respondent driven sampling (RDS, \cite{Heckathorn1997a}) as part of a biological and behavioral surveillance survey in 2011\citep{JohnstonMauritius}.  In that survey, PWID were defined as males or females, of at least fifteen years in age who injected drugs in the previous three months and were living in Mauritius.  In this paper we estimate the HIV and Hepatitis C prevalence, and the proportion of PWID in Mauritius who are female.  As the data were collected using RDS, we next describe the RDS recruitment process, as well as the accompanying estimation methods and the assumptions that they require to produce valid inference.      

%\note{put paragraph about how the study that gave rise to these data}

\subsection{Respondent Driven Sampling Background}

RDS is a network sampling method typically used to infer population proportions of binary traits in hard-to-reach human populations.  RDS has been widely adopted to estimate the prevalence of disease or risk behaviors within high-risk hard-to-reach human populations, including PWID, sex workers, and men who have sex with men. 
 It has been used in hundreds of studies around the world \citep{Johnston2008,Johnston2016,Montealegre2013}, for surveys of biological behavioral surveillance funded by the Global Fund to monitor HIV prevalence, assess risk and program coverage and to measure trends over time \citep{Lansky2007}.  Despite its wide use in settings of public health importance, the statistical properties and optimal inferential strategies for data resulting from RDS still require much additional study.

%MQO:  this paragraph seems disorganized to me.  I started re-arranging it, but it got convoluted so instead I leave you with the suggestion to re-organize.
In RDS, the sampling is a variant of a link-tracing network sampling procedure \citep{HandcockComment2011}.   Link-tracing sampling has been widely used in hard-to-reach populations \citep{Goodman1961Snowball,SHTR1997,sheil1968community}.  In link-tracing, a number of individuals from the target population are enrolled into the study as `seeds', and subsequent samples are selected based on their network connections with previous sample members. 

Networks are used to represent systems of inter-related entities.  In social networks, people (or groups of people) are represented by nodes, and their inter-relationships are represented by edges. \citet{Frank1977} presented an overview of network concepts.  Critically, for our work, %, which we summarize here. 
edges may be either directed (relationships may or may not be reciprocated) %which implies that the relationship has the potential to not be reciprocated, 
or undirected (every relationship is reciprocated). Two nodes are considered incident to each other if they are connected by an edge. RDS draws its name from the fact that respondents are responsible for recruitment by distributing uniquely identified coupons to population members known to them, who are then asked to enroll those they know into the sample, and so on.  

Because the sampling process depends on the network structure \citep{Crawford2014,Verdery2015}, the sample mean (or proportion) from a link-tracing sample is typically a biased estimator of the population mean (or proportion).  Staying within the design-based frame, existing RDS prevalence estimators utilize estimates of sampling probabilities $\pi$, which are typically a function of a respondent's reported number of social ties in a population, called their {\it degrees.}  It should be noted that there are many possible ways of defining the sampling probabilities $\pi$.  Here we define $\pi$ as marginal without-replacement sampling probabilities, marginal over all selections of seeds.  In practice, RDS diverges from its theoretical approximations in several ways, as discussed in \citet{Gile2010Assessment, LuRDS2012,Goel2010,gile2011improved,TomasGile,Lu2013Directed,Rocha2016,Aronow2015}.

\subsection{Edge Inclusion Probabilities}
Social networks tend to have complex structure, and are often difficult to observe in their entirety.  
Typical mechanisms for sampling networks can either rely on pre-specified global rules determining sampling probabilities (i.e. simple random sampling on nodes), or rely on local decision procedures for growing a sample, such as by tracing network edges from previously-observed nodes.  
Examples of the latter include snowball sampling \citep{Goodman1961Snowball,HandcockComment2011}, adaptive web sampling  \citep{ThompsonAdaptiveWeb}, targeted random walk sampling \citep{thompson2006a}, Bayesian adaptive link tracing \citep{StClair2011,ChowThompson2003}, and RDS \citep{Heckathorn1997a,Heckathorn2002II,Salganik2004,VolzHeck2008}.  In each of these network sampling strategies, initial nodes are chosen in some fashion, and then some subset of the nodes incident to the initial nodes are sampled.   This procedure of sampled nodes recruiting a certain number of their neighbors is repeated until the desired sample size is achieved.  In this way, both nodes and edges are observed.  Note that in typical RDS practice, only edges traversed by the sampling process are observed.

Often these network sampling approaches build on the theory of random-walks \citep{Lovasz1993}, where the random walk forms a first-order Markov chain on the space of nodes \citep{Goel2009}.  Less-commonly considered is the related implied distribution on traversed network edges.  Consider an idealized random walk of the following form:   
\begin{enumerate}
\item The network is undirected and connected, (i.e. consisting of a single connected component), without self-ties (loops)
\item An initial node is selected with probability proportional to degree:  $p_1(i) = \frac{d_i}{\sum_{i=1}^N d_i}$
\item Subsequent nodes are selected completely at random, with replacement from among the contacts of the prior sampled node:  $P(S_{k+1}=j)=\left\{ \begin{array}{ll} \frac{1}{d_{S_{k}}} & Y_{S_{k}j} = 1\\ 0 & \textrm{else} \end{array} \right.$,
\end{enumerate}

where $N$ is the population size, $d_i$ is the degree of node $i$, $p_k(i)$ is the probability of sampling the $i^{th}$ node at the $k^{th}$ step, $S_k$ is the index of the node sampled at the $k^{th}$ step, and the $N \times N$ matrix $\bf Y$ represents the sociomatrix of network ties, such that $Y_{ij}=Y_{ji}=1$ if there is an edge between $i$ and $j$, and $Y_{ij}=Y_{ji}=0$ otherwise.  Then the draw-wise edge sampling probabilities are uniform \citep{Salganik2004, Ottprobs}.  Several methods for RDS data, including the estimator in \cite{Salganik2004} rely on treating edge sampling probabilities as equal.  However, \cite{Ottprobs},  show that in without-replacement link-tracing sampling, such as RDS, edge sampling probabilities are not uniform.

Our purpose in this paper is to find the proportion of PWID in Mauritius who are HIV-positive, Hepatitis-C positive, and who are female with a new estimator that improves upon the RDS estimator in \cite{Salganik2004}, (also referred to as the SH, or RDS-I estimator) by adjusting for the bias induced by without-replacement sampling in both the estimation of average degree and accounting for non-uniform edge inclusion probabilities.  In Section \ref{prevest} we introduce the most commonly used RDS prevalence estimators and explain how these estimators are subject to bias when there are non-uniform edge inclusion probabilities.  In Section \ref{estedge}, we propose a method for estimating edge inclusion probabilities, and in Section \ref{improve} we address limitations in current methodology by presenting a prevalence estimator which utilizes estimated edge inclusion probabilities that is particularly suited to the Mauritius data.  In Section \ref{sim} we compare this new RDS prevalence estimator to existing estimators through simulation studies.  In Section \ref{app} we apply this novel method to PWID in Mauritius and estimate the prevalence of HIV and hepatitis C, as well as the proportion of PWID who are female. In Section \ref{dis} we present a brief discussion.

\section{RDS prevalence estimators}\label{prevest}

Most RDS inference is aimed at estimating the population proportion of a binary covariate, or the population prevalence:
\begin{equation}
\label{muparam}
\mu = \frac{1}{N}\sum_{k=1}^N z_k,
\end{equation}
where $N$ is the total population size, and $z_i$ is a binary quantity of interest for the $i^{th}$ unit.  For example, in the Mauritius data, which motivates this paper, the binary covariates are HIV positive status, Hepatitis C positive status, and female gender. Several RDS prevalence estimators have been proposed and implemented \citep{Heckathorn1997a,Heckathorn2002II,Salganik2004,VolzHeck2008,gile2011improved,GileModel2015}, however it has been demonstrated that no one estimator is superior in all cases \citep{TomasGile}. We briefly review the three most commonly implemented estimators of $\mu$: the Volz-Heckathorn estimator (VH), the successive sampling estimator (SS), and the Salganik-Heckathorn estimator (SH), and their relationships to the proposed estimator \citep{Johnston2016}. The VH and SS estimators estimate the probability of observing individuals based on their degree, and use this probability to perform inverse-probability weighting.  The SH estimator relies on the assumption that each edge has an equal probability of being included in the sample, and leverages edge-wise information to estimate prevalence.  The SH out-performs the alternatives in the presence of homophily, the tendency for individuals with similar attributes to be connected with each other, and differential recruitment effectiveness, that is, when the target populations forms ties preferentially among similar people, and when one group tends to have more successful recruitments per recruiter \citep{TomasGile}.  It also performs especially well when the initial sample is highly unrepresentative of the overall population, such as when all individuals selected in the initial sample are HIV positive \citep{GileModel2015}.  However, it can be severely biased when the sample fraction is large \citep{Gile2010Assessment}.  Note that the the improvement offered by the SS over the VH is the adjustment for without-replacement sampling.  In this paper, we use an approximation to the sampling process similar to that in the SS to create a new estimator similar to the SH, but adjusting for without-replacement sampling.  Other methods follow the RDS sampling procedure, but then collect additional information about each sampled individual's ego network \citep{lu2013linked} which is then utilized in the estimation process.  In this work we focus on how to improve RDS estimation without collecting additional information. First, we describe the VH, SS, and SH estimators.

The VH estimator \citep{VolzHeck2008} assumes that the RDS sample can be treated as an independent sample from the stationary distribution of a random walk on the space of network nodes.  Because the stationary distribution is proportional to the nodal degrees ($d_i$'s), this estimator inverse-weights the observed nodal values of the quantity of interest ($z_i$'s), by the degrees, in a generalized Horvitz-Thompson estimator \citep{Thompson2002} ratio format:

\begin{equation*}
\label{muVH}
\widehat{\mu_{VH}} =\frac{\sum_{i=1}^n \frac{z_i}{d_i}}{\sum_{i=1}^n \frac{1}{d_i}},
\end{equation*}

\noindent where $n$ is the sample size and nodes are ordered such that the sampled nodes appear first.

While VH performs well in many settings it is subject to bias under several sampling conditions including differential recruitment effectiveness (individuals passing coupons in one group are likely to disperse more of their coupons than individuals in the other group), and differential activity (individuals in one group tend to have a higher degree than individuals in the other group) in the presence of a large sample fraction\citep{TomasGile}.  

The SS estimator \citep{gile2011improved} has a form very similar to the VH estimator.  While the VH estimator assumes that sampling probabilities are proportional to degree, the SS estimator approximates these probabilities based on a without replacement process  \citep{gile2011improved}:

\begin{equation*}
\label{muSS}
\widehat{\mu_{SS}} =\frac{\sum_{i=1}^n \frac{z_i}{\hat\pi_i({\bf{d}})}}{\sum_{i=1}^n \frac{1}{\hat\pi_i({\bf{d}})}}.
\end{equation*}

\noindent The formulation of the SS estimator differs from the formula for the VH estimator as the estimated sampling probability of $i$ is a function of both $d_i$, as well as the degree sequence of the entire sample (noted as ${\bf{d}}$).  The SS estimator is not subject to bias when there are large sampling proportions, though it is still subject to bias resulting from other conditions including differential recruitment effectiveness in combination with homophily.  Its finite population correction also relies on a working estimate of the population size.  

In contrast to the VH and SS estimators, the SH estimator \citep{Salganik2004} relies heavily on the number of within and between group edges (i.e. recruitments) in an RDS sample, rather than a weighted proportion of the sample that has the attribute of interest.  For this reason, the SH estimator is especially sensitive to unequal edge sampling probabilities.  For example, if the RDS recruitment process has a disproportionate number of recruitments from someone who has HIV to someone who does not have HIV (relative to the number of social ties between people in these two groups) the SH estimator will be heavily biased towards underestimating the proportion of people with HIV.  However, it is also because of this very different formulation that uses the number of between group edges that the SH performs well in circumstances where the alternative VH and SS perform poorly, in particular in the face of the combination of differential recruitment effectiveness, differential activity, and homophily.  %, rather than the number of nodes from the group of interest.  
Because our proposed estimator builds on the SH, we describe its form in greater detail.

While we assume that the underlying network is undirected, in the RDS sample we observe edges as they are traversed in a directed manner, so we treat equivalence classes of observed directed edges.  For instance, consider the case where we are interested estimating the proportion $\mu$ of a networked-population that is HIV positive, also equal to the population average of a binary nodal variable $z_i \in \{0,1\}$, where $z_i=1$ if node $i$ is infected.  Define
\begin{equation*}
\label{T}
T_{(k,1-k)}= \frac{\sum_{i:z_i=k} \sum_{j: z_j = 1-k}  Y_{ij}}%, j \text{ neighbor of } i)}
{N_k }, 
\end{equation*}
to be the average number of ties a single type $z_i=k$ node has to type $z_i=1-k$ nodes, $k \in \{0,1\}$, where $N_k$ is the population number of nodes of type $k$.  Since the network is undirected, $N_0 T_{(0,1)} = N_1 T_{(1,0)}$, so $\frac{T_{(1,0)}}{T_{(0,1)}} = \frac{N_0}{N_1}$, and  
\begin{equation*}
\label{muT}
\mu =\frac{N_1}{N_0+N_1}=\frac{ T_{(0,1)}} {T_{(0,1)} + T_{(1,0)}}.
\end{equation*}

We can express $T_{(k,1-k)}$ in terms of $D_{(k)}$  and $C_{(k,1-k)}$, where 
$D_{(k)} = \frac{1}{N_k} \sum_{i:z_i=k} d_i$ is the average degree for those nodes of type $k$ and 

\begin{equation*}
\label{Ck1k}
C_{(k,1-k)} = \frac{\sum_{i:z_i=k} \sum_{j: z_j = 1-k}  Y_{ij}}{\sum_{i:z_i=k} d_i}
\end{equation*}

\noindent is the proportion of cross-group ties among the ties of type $k$ nodes.  Then the proportion of proportion with the characteristic of interest is calculated as:

\begin{equation}
\label{muDC}
 \mu =\frac{D_{(0)} C_{(0,1)}}{D_{(0)}  C_{(0,1)}+D_{(1)}  C_{(1,0)}}.
\end{equation}

%MQO:  is there a clause missing from the below sentence?
\noindent The \citet{Salganik2004} method takes the form of the above equation and utilizes the observed between and within group referrals and the degree for each participant sampled through RDS.  They estimate:

\begin{equation}
\label{estc}
\widehat{C}_{(k,1-k)} = \frac{r_{(k,1-k)}}{r_{(k,1-k)}+r_{(k,k)}}, ~~~ k \in \{0,1\},
\end{equation}

\noindent where $r_{(k,1-k)}$ is the number of referrals from an $k$ node to a $1-k$ node, $k, \in \{0,1\}$.  This is based on the assumption that the sampling process can be treated as from the stationary distribution of a Markov chain on the network nodes, leading to uniform edge-traversal probabilities.  
$D_{(1)}$ and $D_{(0)}$ are also unknown and need to be estimated in order to compute prevalence estimates. Like the VH estimator, the SH estimator makes use of the generalized Hansen-Hurwitz estimator \citep{HansenHurwitz1943, Thompson2002}, but rather than estimating the prevalence, the SH estimator uses this form to estimate the average degree for each group: 

\begin{equation}
\label{estgroupdegrees}
\begin{aligned}
\widehat{D}_{(1)} & =\frac{\sum_{i=1}^n d_i z_i}{\sum_{i=1}^n \frac{z_i}{d_i}},\text{   }
\widehat{D}_{(0)} & =\frac{\sum_{i=1}^n d_i (1-z_i)}{\sum_{i=1}^n \frac{1-z_i}{d_i}}.
\end{aligned}
\end{equation}

Substituting these estimates directly into (\ref{muDC}) gives the form of the SH estimator:

\begin{equation}
\label{muSHex}
\widehat{\mu_{SH}} =\frac{\widehat{D}_{(0)}  \widehat{C}_{(0,1)}}{\widehat{D}_{(0)}\widehat{C}_{(0,1)}+\widehat{D}_{(1)} \widehat{C}_{(1,0)}}.
\end{equation}

The SH estimator out-performs other estimators in certain situations, particularly when there is differential recruitment effectiveness \citep{TomasGile}.  However, the SH estimator has been noted to perform poorly in the presence of differential activity, and when there is a large sample fraction \citep{Gile2010Assessment}.

\subsection{Implication for Inference on Nodal Characteristics: Salganik-Heckathorn Estimator}

In equation \ref{muSHex} we see that the SH estimator relies on intermediate estimates (\ref{estc}) and (\ref{estgroupdegrees}) to estimate $\mu$.  While the form of (\ref{estgroupdegrees}) is based on weighting the nodal sample and does not depend directly on assumptions about edge sampling probabilities, (\ref{estc}) relies heavily on the assumption of uniform edge sampling probabilities.  When these probabilities are far from uniform, as in without-replacement sampling with substantially large sample fractions, the estimation in (\ref{estc}) may be quite inaccurate, leading to biased estimates produced by the SH estimator.  These biases will be exacerbated in the presence of differential activity.  Based on the analysis in \cite{Ottprobs}, we expect that $\widehat{C}_{(0,1)}$  will have positive bias, and $\widehat{C}_{(1,0)}$ will have negative bias when group $0$ has higher mean degree than group $1$, leading to positive bias in $ \widehat{\mu_{SH}}$.

\section{Estimating Edge Inclusion Probabilities}\label{estedge}

The without replacement sampling design in RDS results in unequal edge sampling probabilities which have previously been unaccounted for \citep{Ottprobs}.  In order to address the estimation problem resulting from unequal edge inclusion probabilities, we propose to estimate edge inclusion probabilities and use inverse-probability weighting with these weights to improve the SH estimator.    

Because of the complexity of the RDS sampling process, we must estimate these probabilities under an approximation to the sampling process.  In particular, we follow \cite{gile2011improved} and use a successive sampling approximation to the sampling process.  Successive sampling, also known as probability proportional to size without replacement (PPSWOR) sampling, is a sampling mechanism used to draw a sample of size $n$ from a population of size $N$ of units with {\it unit sizes} ${\bf u} = (u_i), ~ i \in \{1 \dots N\}$ \citep{Raj1956,Rao1991,gile2011improved}.  It proceeds as follows:

\begin{enumerate}
\item The first unit is sampled with probability proportional to ${\bf u}$.
\item Each subsequent sample is drawn with probability proportional to ${\bf u}$ {\it from the previously unsampled units}, resulting in step-wise sampling probabilities:
\begin{equation}
P(S_k=i|S_1, \ldots , S_{k-1}) = \left\{ \begin{array}{ll}  \\
\frac{u_i}{\sum_{j=1}^N u_j - \sum_{j=1}^{k-1} u_{S_j}} & i\notin \{S_1, \ldots , S_{k-1}\} \\
0 & \textrm{else}.
\end{array} \right.
\label{SSprobs} 
\end{equation}
\item Sampling ends when sample size $n$ is attained.
\end{enumerate}

\cite{gile2011improved} notes that treating nodal degrees ${\bf d}$ as unit sizes (${\bf u}$ above), the probabilities (\ref{SSprobs}) are equivalent to the step-wise sampling probabilities of a without-replacement random walk on a network drawn from a Molloy-Reed distribution \citep{MolloyReed} conditional on the population degree distribution $\mathbb{N}=\mathbb{N}_1,\mathbb{N}_{2},\ldots, \mathbb{N}_K$ where $\mathbb{N}_j$ is the number of nodes with degree $j$, and $K$ is the maximum degree.  Recall that $\bf{d}$ is the vector of degrees in the sample, whereas $\mathbb{N}$ is the distribution of degrees in the population.  %This argument also implies that directed edge sampling probabilities from a successive sampling process are equivalent to the directed edge sampling probabilities from this Molloy-Reed approximation.  

The approximation in \cite{gile2011improved} has proved highly effective in adjusting the VH estimator.  Therefore, it stands to reason that using the corresponding directed edge weights to adjust the SH estimator will be similarly effective in accounting for finite population effects.  We therefore present an approach to estimating directed edge sampling weights based on a successive sampling approximation to the RDS process.  

Assuming that there is an underlying network which we cannot fully observe, we will estimate the probability that we observe in the RDS sample the directed edge between node $i$ and node $j$, given that $i$ and $j$ are connected in the underlying graph: $P(i\rightarrow j)$.
 
We specify that $S_{i,j} =1 $ if node $i$ is sampled as a non-terminal node while node $j$ is still unsampled. Then:
\begin{equation}
\label{eprob1}
P(i\rightarrow j)= P(i\rightarrow j, S_{i,j}=1)=P(i \rightarrow j| S_{i,j}=1)P(S_{i,j}=1).
\end{equation}

\noindent Because successive sampling acts on equivalence classes of units (nodes) based on nodal degrees, we treat equivalence classes of directed edges based on ordered pairs of nodal degrees.  We make the approximation: 

\begin{equation}
\label{eprob2}
P(i \rightarrow j| S_{i,j}=1)\approx \min \left(\frac{n_c}{h(d_i)}, 1 \right)\approx \min \left(\frac{n_c}{g(d_i)}, 1 \right),
\end{equation}

\noindent where $h(d)$ is the average number of connections incident to a node with degree $d$ that are unsampled when such a node is sampled, and, $n_c$ is the maximum number of coupons that each participant may pass on.  We further approximate $h(d)$ as $g(d)= d(1-\mu(d)/N)$, where $\mu(d)$ is the average number of nodes that have been sampled when a node with degree $d$ is sampled.  The intuition for this is that if a node $i$ is chosen as a seed, then $\mu(d_i)=0$ and $P(i \rightarrow j| S_{i,j}=1)= n_c/d_i$.  As the sampling process continues, more nodes are included, and thus are not available to be sampled again.  If a node $i$ is chosen after 10\% of nodes from the population are included in the sample, then we approximate the number of unsampled nodes that $i$ could then recruit into the sample as $0.9d_i$, and $P(i \rightarrow j| S_{i,j}=1)= n_c/(0.9d_i )$. Here since we are marginalizing over all nodes with degree $d_i$, we find the average sampling order of nodes with degree $d_i$ as $\mu(d_i)$. We use these approximations so that we can efficiently estimate them from a PPSWOR sample without instantiating an unknown network.  

The more challenging aspects of estimating $P(i\rightarrow j)$ are estimating $P(S_{i,j}=1)$ and $\mu(d)$.   \citet{gile2011improved} faces a similar challenge of estimating nodal inclusion probabilities in RDS in the presence of large sample fractions.  In \citet{gile2011improved}'s successive sampling approximation, the nodal inclusion probabilities are not available in closed form, and must be estimated by an iterated simulated sampling process.  This procedure converges to sampling probabilities $\hat{\pi}_k =\mathrm{f}(k,n,\mathbb{N})$, for equivalence classes of nodes according to degree $k$, and dependent on sample size $n$ and population distribution of degrees, $\mathbb{N}$.

Following \cite{gile2011improved}, we estimate the nodal sampling probabilities $\mathrm{f}(k,n,\mathbb{N})$ and the degree distribution $\mathbb{N}$ with successive sampling.  We then conduct an additional round of simulated resampling to estimate $P(S_{i,j}=1)$ and $\mu(d)$.  Given the estimated degree distribution $\mathbb{N}$, we simulate $M$ resamplings according to the following procedure:

\begin{enumerate} 
  \item For $t$ in $1 \ldots M$:
  \begin{enumerate}
      \item Draw a sample $S_1, \ldots, S_n$ of size $n$ from $\mathbb{N}$ using the successive sampling method treating nodal degrees as unit sizes, with initial node chosen with probability proportional to degree and subsequent nodes drawn according to (\ref{SSprobs}).
      \item For all ordered pairs $k,l$ such that $k, l \in 1,\ldots, K$, record the number of times a node with degree $k$ is sampled before a node with degree $l$ forming a $K \times K$ matrix $V^t$ (where $K$ is the largest degree in the degree distribution).  For example, suppose that $\mathbb{N}_4=3$ and $\mathbb{N}_5=2$, and the simulated sample includes two nodes of degree $4$ and one of degree $5$ in the order $4,4,5$. %Suppose that in our simulated $t$ ordered sample, that first a node with degree $i$ is sampled, followed by a node with degree $i$, and lastly, a node with degree $j$.  
      Then $V^t_{44}=3, V^t_{55}=1, V^t_{45}=4, V^t_{54}=1$, all other entries of $V^t$ are zero, and we would record this in a $5\times 5$ matrix $V^t$:\\
\[
\begin{bmatrix}
    0 & 0 & 0& 0 & 0  \\
    0 & 0 & 0& 0 & 0   \\
    0 & 0 & 0& 0 & 0   \\
    0 & 0 & 0& 3 & 4   \\
    0 & 0 & 0& 1 & 1  
\end{bmatrix}
\]   
      \item Let $\hat{\mu}^t(k)$ be the average ordered index of all nodes of degree $k$ observed in the simulated sample, and ${g^t}(k)=k(1-\frac{\hat{\mu}^t(k)}{N})$.  Treat $g^t(k)$ as null whenever no such nodes are sampled.
  \end{enumerate}
%  \item Repeat steps 1 and 2 $M$ times, resulting in $M$ matrices: $V^1, \ldots, V^M$.
  \item Let $W$ be a $K \times K$ matrix in which $W_{ij}$ is the proportion of instances in the simulation in which a node with degree $i$ was sampled before a node with degree $j$ where:
  \[ W_{kl}=\frac{\sum_{t=1}^MV^t_{kl}+1}{\mathbb{N}_k\times 
 \mathbb{N}_l \times M +1}, ~~~ k\neq l \textrm{, and } W_{kk}=\frac{\sum_{t=1}^MV^t_{kk}+1}{(\mathbb{N}_k-1)\times \mathbb{N}_k \times M +1}, ~~~ k \in 1, \ldots, K.
\]
 \item Estimate $P(S_{i,j}=1)$  with $W_{d_id_j}$, for  $i,j \in 1, \ldots N$.  
 \item Estimate $\hat{g}(k)$ as the average of the non-null values of the $M$ realizations of $g^t(k)$, for $k \in 1 \ldots K$. 
%  \item For the diagonal elements of $W$: $W_{ii}=\frac{\sum_{t=1}^MV^t_{ii}+1}{(\mathbb{N}_i-1)\times \mathbb{N}_i \times M +1}$
\item Estimate $P(i \rightarrow j)$ for equivalence classes of $d_i$ and $d_j$ as:
 \[
%P(i \rightarrow j)= 
\hat{q}_{d_i, d_j} = \min \left(\frac{n_c}{\hat{g}(d_i)}, 1 \right) W_{d_id_j}.
\]
\end{enumerate}

\section{The weighted SH estimator}\label{improve}

Recall that SH estimates prevalence by estimating the average number of cross-ties from each group to the other group.  %someone who is HIV negative to HIV positive, and the average number of cross-ties from someone who is HIV positive to someone who is HIV negative.  
Here we use the new estimated inclusion probabilities ($\hat{q}_{d_i,d_j}$) to improve upon the SH estimator by weighting observed edges and nodes.  We refer to this new estimator as the {\it Weighted SH Estimator}.  Let $r_{i,j}$ be the indicator that person $i$ passed a coupon to person $j$, and recall that $z_i=1$ if $i$ is HIV positive, and $z_i=0$ if person $i$ is HIV negative. Then we have:

\begin{equation*}
\label{Cest10}
\widehat{C}_{W(1,0)} = \frac{\sum_{i:z_i=1} \sum_{j:z_j=0}r_{ij}/\hat{q}_{d_i,d_j}}{\sum_{i:z_i=1}\sum_{j:z_j=0}r_{ij}/\hat{q}_{d_i,d_j}+\sum_{i:z_i=1} \sum_{j:z_j=1, i\neq j}r_{ij}/\hat{q}_{d_i,d_j}} ,    
\end{equation*}

\begin{equation*}
\label{Cest01}
\widehat{C}_{W(0,1)} = \frac{\sum_{i:z_i=0} \sum_{j:z_j=1}r_{ij}/\hat{q}_{d_i,d_j}}{\sum_{i:z_i=0}\sum_{j:z_j=1}r_{ij} /\hat{q}_{d_i,d_j}+\sum_{i:z_i=0} \sum_{j:z_j=0, i\neq j}r_{ij}/\hat{q}_{d_i,d_j}} .  
\end{equation*}

The SH estimator relies on the assumption that individuals are sampled in proportion to their degree in order to estimate the average degree for each group.  %those who are HIV positive and those who are HIV negative.  
Here we follow \citet{gile2011improved} and make use of the SS estimator to estimate $D_{(1)}$ and $D_{(0)}$ using $\hat{\pi}_{d_i}$, the estimated probability that someone with degree $d_i$ is included in the sample:

\begin{equation*}
\label{DWest1}
\widehat{D}_{W(1)}=\frac{\sum_{i} z_i d_i{\hat{\pi}_{d_i}}^{-1}}{\sum_{i}z_i{\hat{\pi}_{d_i}}^{-1}},
\end{equation*}

\begin{equation*}
\label{DWest0}
\widehat{D}_{W(0)}=\frac{\sum_{i}(1-z_i)d_i{\hat{\pi}_{d_i}}^{-1}}{\sum_{i} (1-z_i){\hat{\pi}_{d_i}}^{-1}}.
\end{equation*}

\noindent Now we have 

\begin{equation}
\label{Pest1}
 \widehat{\mu}_{WSH} =\frac{\widehat{D}_{W(0)} \widehat{C}_{W(0,1)}}{\widehat{D}_{W(0)} \widehat{C}_{W(0,1)}+\widehat{D}_{W(1)} \widehat{C}_{W(1,0)}},
\end{equation}

\noindent which we use to estimate prevalence.  Note that the weighted SH prevalence estimator is in the same form as the original SH estimator, but includes adjustments for unequal edge sampling probabilities.

\subsection{Variance Estimation of Weighted SH Estimator}

%\boxit{I'm not sure what needs to go here}

Uncertainty estimation for RDS estimators is typically conducted using a bootstrap procedure, most often the procedure introduced in  \citet{Salganik2006}.  Little is known about the performance of this bootstrap, although the studies that do exist suggest that it is anti-conservative 
%Variance estimation of RDS prevalence estimators is an ongoing area of research.   Since RDS may have increased variance due to the inherently correlated observations, the variance for an RDS prevalence estimate will in general be larger than a mean taken from a simple random sample. Currently the standard variance estimator for RDS prevalence estimates is the bootstrap variance estimator put forth by \citet{Salganik2006}.  
%However, the Salganik bootstrap variance estimate for RDS point estimators tends to be under-estimate the variance 
\citep{Goel2010,Nesterko2015,Wejnert2009,Verdery2015Variance}.  %knote:  add cite to CDC paper later.  
Creating an uncertainty estimator that corrects the under-estimation of the Salganik bootstrap is a separate line of inquiry beyond the scope of this project.  We therefore propose to apply a version of the Salganik bootstrap, and suggest that users remember that this estimate should be regarded with all the caveats applied to other RDS uncertainty estimators.  The resulting bootstrap estimation procedure proceeds as follows:
%Understanding the variance of the weighted SH estimator that we propose here, as well as the variance of other RDS estimators, is important though we do not address it in this work.  Here we use the Salganik bootstrap variance estimate, which we describe here:
\begin{enumerate}
  \item Categorize nodes in the model as either referred by infected, or referred by uninfected.
  \item Sample uniformly and with replacement from the RDS sample $n_s$ seeds, where $n_s=$ the number of seeds in the RDS sample.  These become our bootstrap seeds.
  \item For each new member in our bootstrap sample that is infected, sample $n_c$ with replacement from the observed RDS sample that were referred by someone who is infected.
  \item For each new member in our bootstrap sample that is uninfected, sample $n_c$ with replacement from the observed RDS sample that were referred by someone who is uninfected.
  \item From steps 3 and 4, we have now collected our new wave of sample collection.  Repeat steps 3 and 4 with the newest wave, until the sample size in the original RDS sample is reached.
  \item Calculate the RDS estimate of disease prevalence $\hat{P}_{bs}$.
  \item Repeat steps 2-6 many times to estimate the bootstrap distribution of $\hat{P}_{bs}$.
  \item Use the distribution of $\hat{P}_{bs}$ to form confidence intervals or calculate standard errors using a Normal approximation.
\end{enumerate}  

\noindent In this way we seek to estimate the variance of the the RDS estimate of the population proportion of the characteristic(s) of interest, explicitly, the variance of an RDS estimator if repeated RDS samples of the same sample size, with the same number of seeds, using the same maximum number of coupons, were collected from the same population.   

%\noindent In this way we seek to estimate the variance of the the RDS estimate of the population proportion of the characteristic(s) of interest. Taking the Mauritius data as an example, we seek to estimate the variance of an RDS estimator if repeated RDS samples of the same sample size, with the same number of seeds, using the same maximum number of coupons, were collected from the same population of PWID in Mauritius.   

\section {Simulation Studies}\label{sim}
The goal of this project is to create an improved estimator for RDS to better estimate the HIV prevalence, Hepatitis C prevalence, and proportion female of PWID in Mauritius.  To do so, we propose an inferential method relying on estimation of edge-sampling probabilities.  Here, we present simulations to illustrate the performance of our proposed RDS estimator.

\subsection{Simulations to compare RDS prevalence estimators} 
To evaluate the performance of the proposed estimator, 
we simulated RDS on networks and compared five prevalence estimators:  the mean of the sample, the SS, the VH, the original SH, and the weighted SH. To allow for comparisons with previous RDS work, we used simulated networks that were used in \citet{TomasGile} and \citet{Gile2010Assessment}, where a detailed description of the methods used to generate these networks can be found.  These networks are each composed of 1000 nodes composed of two groups: {\emph{infected}} (200 nodes) and {\emph {uninfected}} (800 nodes).  In the networks used in the simulations, the networks either have a homophily value of one or two and differential activity of one or two.  Additionally, as we simulate RDS on these different networks, we vary the sampling proportion (20\%, 50\%, 70\%), and the differential recruitment effectiveness (1 and 1, or .9 and .6).  These terms are defined mathematically in Table~\ref{tab:NetworkParams} for a network $Y$ with $N$ nodes where $N^1$ is the number of infected nodes in the population and $N^0=N-N^1$.  We use these parameters to determine the parameters of an exponential random graph model, then sample networks from these models using the {\tt statnet} R package \citep{statnet}.  The parameters network statistics specified in each condition are then the expected values under the network-generating model.

\singlespacing
\begin{table}
\caption{Network Parameters for Simulations}
\label{tab:NetworkParams}
\begin{center}
\begin{tabular}{l || c}
%\multicolumn{2}{c}{{\bf Table 4.4.} Network Parameters for Simulations} \\ \hline \hline
 Parameter & Definition \\ \hline  \hline 
Number of Nodes &  $N$ \\
\phantom{ \tiny }\\
Prevalence & $\mu = \sum_{i=1}^N z_i /N$\\
\phantom{  }\\
Mean degree & $\bar{d} = 1/N \sum_{i=1}^N d_i/2 $\\ 
\phantom{  }\\
Homophily & $H=\frac{2/(N^1(N^1-1)) \sum_{i,j}z_iz_j\mathbb{E}(y_{ij})}{1/(N^1N^0) \sum_{i,j}z_i(1-z_j)\mathbb{E}(y_{ij})}$\\
\phantom{  }\\

Differential Activity & $DA= \bar{d}^1/\bar{d}^0 $\\
\hline
\end{tabular}
\end{center}

\end{table}
\doublespacing

Differential recruitment effectiveness occurs when individuals in one group are more likely to successfully recruit people into the sample.  In these simulations differential recruitment effectiveness was set to either (1,1) or (.9,.6).  Differential recruitment effectiveness = (1,1) when individuals in both the infected and uninfected groups would recruit as many people into the sample as they were allowed.  Differential recruitment effectiveness = (.9, .6) when individuals in the infected group have a 90\% chance of successfully recruiting someone for each coupon they are given, and those in the uninfected group have a 60\% chance of successfully recruiting someone for each coupon they are given.  In these simulations we begin the sampling process with 10 seeds, and the number of coupons, $n_c=2$, so each person in the sample recruited up to two others given the sample size had not yet been attained.

\begin{figure}[tbp] % float placement: (h)ere, page (t)op, page (b)ottom, other (p)age
  \centering
  % file name: C:/Users/Miles/Dropbox/Miles_Krista/Preparing RDS Re-Weighted Paper for Publication/side_by_side_boxplots_no_differential_activity_nohomo_noRE_SS200_newest.pdf
  \includegraphics[bb=12 31 706 404,width=5.67in,height=3.05in,keepaspectratio]{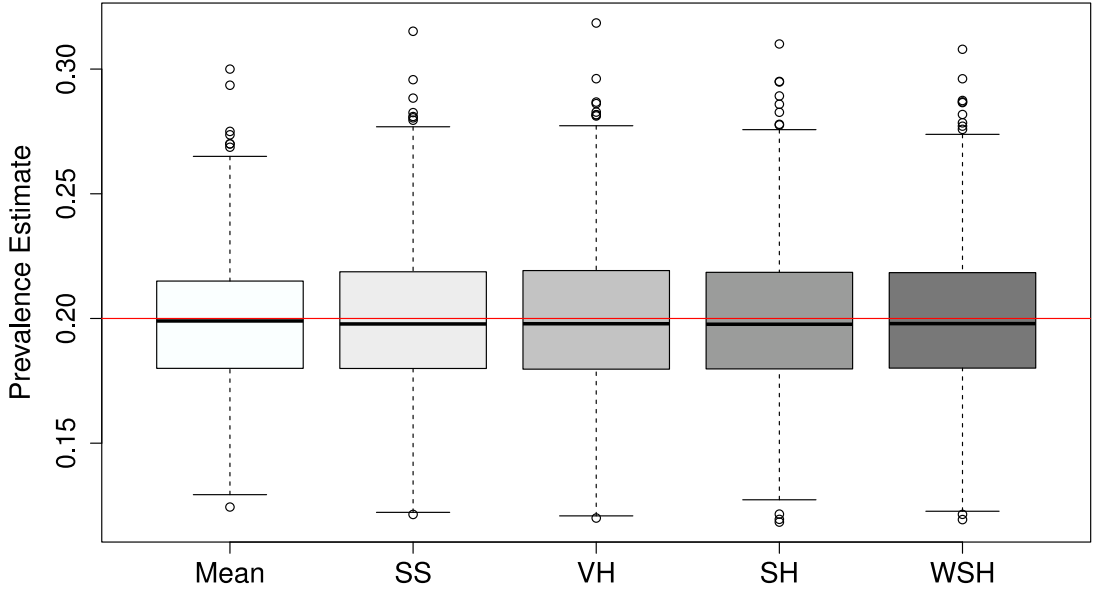}
  \caption{Prevalence estimates using the sample mean, SS, VH, SH, and weighted SH estimators without differential recruitment effectiveness, differential activity, or homophily effects.  Sampling fraction is 20\%. The horizontal red line represents the true prevalence.}
  \label{fig:boxplots_sample_NoHomo_NoRE_NoDA_comparison}
\end{figure}

In the first set of simulations, we demonstrate how the weighted SH estimator compared to the other RDS prevalence estimators in the absence of differential activity, differential recruitment effectiveness, and homophily effects.  We simulated RDS on 1,000 networks of size 1,000, with prevalence 0.20, and a sampling proportion of 20\%. In these networks, the average degree was 7.07. The prevalence estimates from these 1,000 simulations are presented in Figure~\ref{fig:boxplots_sample_NoHomo_NoRE_NoDA_comparison}, and the MSE is displayed in the first panel of Table~\ref{tab:Homo_RE_comparison_all}.  All five estimators, including the naive mean perform comparably when there is no differential activity, differential recruitment effectiveness or homophily present, and the sampling proportion is relatively small.

\begin{figure}[tbp] % float placement: (h)ere, page (t)op, page (b)ottom, other (p)age
  \centering
  % file name: C:/Users/Miles/Dropbox/Miles_Krista/Preparing RDS Re-Weighted Paper for Publication/side_by_side_boxplots_sample_size_comparison_newer.pdf
  \includegraphics[bb=12 13 706 404,width=5.67in,height=3.2in,keepaspectratio]{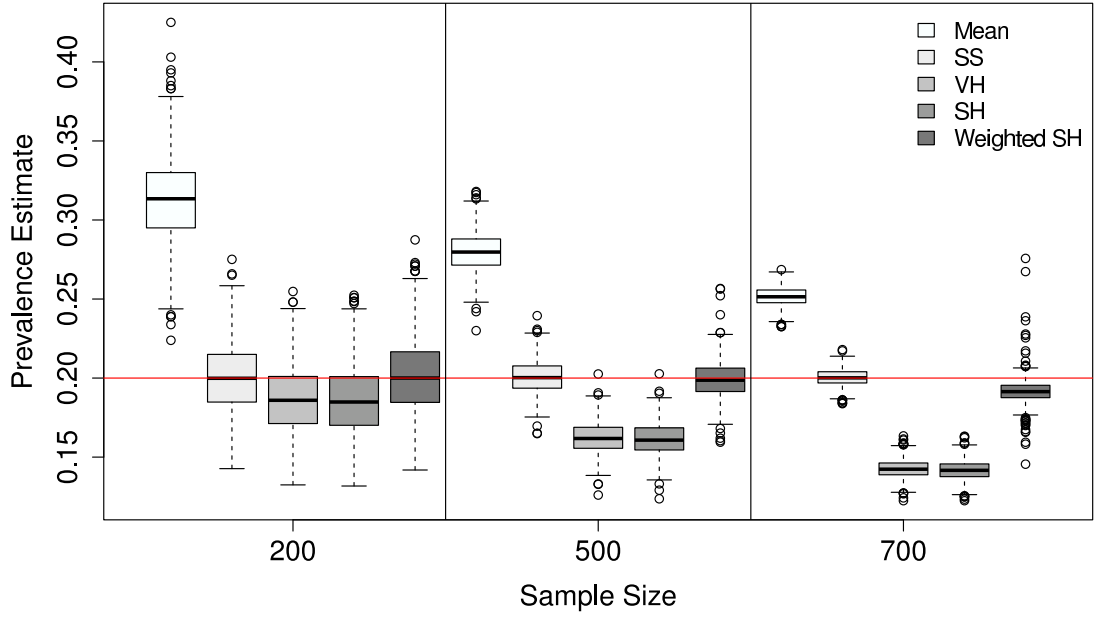}
  \caption{Prevalence estimates using the sample mean, SS, VH, SH, and weighted SH estimators for samples of size 200, 500, and 700 from simulated networks of 1000.  Here differential activity =2, homophily=1, and differential recruitment effectiveness =(1,1).  The horizontal red line represents the true prevalence.}
  \label{fig:side_by_side_boxplots_sample_size_comparison}
\end{figure}

In the second set of simulations, we used 1000 networks where homophily was set to 1 and differential activity was held constant at 2. In the simulated RDS sampling, the differential recruitment effectiveness was set to (1,1), and we varied the sampling proportion (20\%, 50\%, 70\%). In these networks, the average degree for those with the characteristic of interest was 11.7, and 5.83 among those who did not have the characteristic of interest.  Overall, the average degree was 6.98. On each of the 1,000 networks we drew three RDS samples of sizes 200, 500, and 700, to produce the desired sampling proportions.  Boxplots of the estimated prevalence are displayed in Figure~\ref{fig:side_by_side_boxplots_sample_size_comparison}, and the second panel of Table~\ref{tab:Homo_RE_comparison_all} contains the MSE of the various estimators.  In these simulations, regardless of the sampling proportion, the SS estimator performs the best in terms of MSE, and the Weighted SH performing comparably to the SS when the sampling proportion is 20\% or 50\%. Because the VH and SH estimators would perform best when sampling is conducted with replacement, we would expect that larger sampling proportions would incur greater bias for these estimators. Indeed, the VH and SH estimators display increasing bias as the sampling proportion increases, and perform similarly in terms of MSE. The naive mean performs very poorly relative to all the other estimators when the sample size is small, but has a smaller MSE than the VH and the SH estimators when the sampling proportion is 70\%.

\begin{figure}[tbp] % float placement: (h)ere, page (t)op, page (b)ottom, other (p)age
  \centering
  % file name: C:/Users/Miles/Dropbox/Miles_Krista/Preparing RDS Re-Weighted Paper for Publication/side_by_side_boxplots_sample_Homo_RE_comparison_newer.pdf
  \includegraphics[bb=12 13 850 404,width=5.67in,height=2.65in,keepaspectratio]{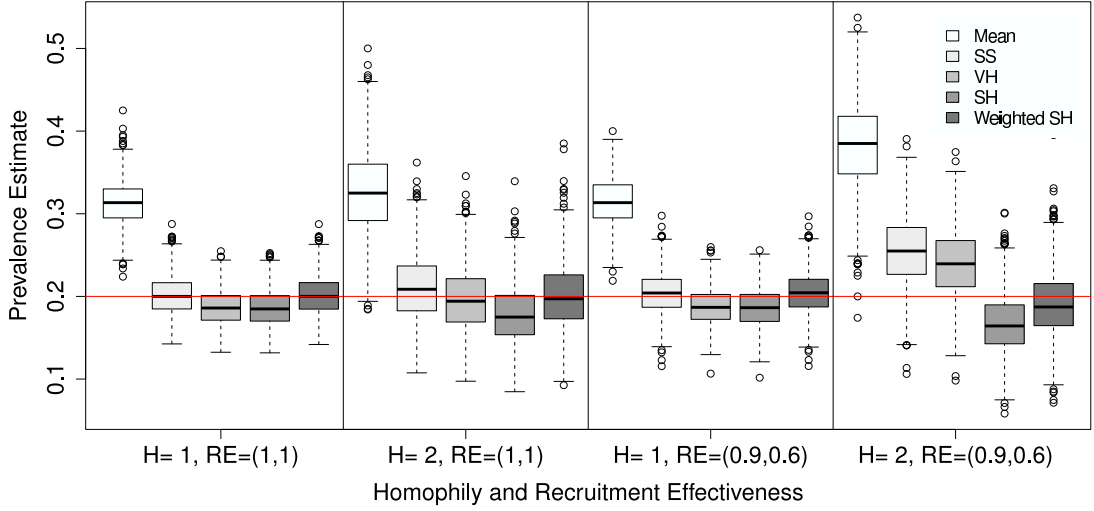}
  \caption{Prevalence estimates using the sample mean, SS, VH, SH, and weighted SH estimators for varying homophily $\in (1,2)$, recruitment effectiveness $\in [(1,1);(.9,.6)]$, and differential activity =2.  Sampling fraction is 20\%.  The horizontal red line represents the true prevalence.}
  \label{fig:side_by_side_boxplots_sample_Homo_RE_comparison}
\end{figure}

The next set of simulations were performed on networks where homophily was varied to be either 1 or 2 and differential activity was held constant at 2.  In each homophily condition, RDS was carried out on each of the networks (1,000 networks where homophily =1, 1,000 networks where homophily=2)
%Among the 1,000 networks where homophily was set to 1, RDS was carried out 
with differential recruitment effectiveness set to (1,1) and (.9,.6).  In these networks, the average degree for those with the characteristic of interest was 11.7, and 5.83 among those who did not have the characteristic of interest.  Overall, the average degree was 6.98.  %Among the 1,000 networks where homophily was set to 2, RDS was carried out with differential recruitment effectiveness set to (1,1) and (.9,.6). 
In this set of simulations, the sampling proportion was held constant at 20\%.  We compare the performance of the estimators from these simulations in Figure~\ref{fig:side_by_side_boxplots_sample_Homo_RE_comparison}, and in third panel of Table~\ref{tab:Homo_RE_comparison_all}.  Regardless of homophily level, when the differential recruitment effectiveness is set to (1,1) all five estimators perform similarly in terms of MSE, except the mean which is positively biased.  Similarly, when homophily is set to 1, and differential recruitment effectiveness is set to (.9,.6) the mean has the highest MSE, and the other four estimators perform comparably. However, when differential recruitment effectiveness is set to (.9,.6) and homophily is set to 2, the weighted SH estimator has the lowest MSE, followed by the SH.  This is consistent with our expectation that the revised estimator corrects for finite population biases (as does the SS), while also maintaining the insensitivity to differential recruitment effectiveness of the SH.  In fact, for all simulation conditions,the weighted SH estimator has lower MSE than the SH.

%\singlespacing
%\begin{table}
%\caption{Simulated Prevalence Estimates on Network with 1000 Nodes, Varying Sample Proportion}
%\label{tab:samplesize}
%\begin{tabular}{c c c c  |c c c |c c c}
%& \multicolumn{3}{c}{MSE $\times 10^4$}&\multicolumn{3}{c}{$|$Bias$|$ $\times 10^3$}&\multicolumn{3}{c}{SE$\times 10^3$}\\  
%  Estimator&\multicolumn{1}{c}{200} &\multicolumn{1}{c}{500}&\multicolumn{1}{c}{700} & \multicolumn{1}{c}{200} & \multicolumn{1}{c}{500}& \multicolumn{1}{c}{700}& \multicolumn{1}{c}{200} & \multicolumn{1}{c}{500} &\multicolumn{1}{c}{700} \\ \hline 
%Mean & 136.54 & 64.67 & 27.22 & 113.66 & 78.47 & 51.81 & 27.14 & 12.25 & 6.13\\
%SS & 4.61 & 1.01 & 0.29 & 0.55 & 0.54 & 0.40 & 21.49 & 10.05 & 5.35 \\
%VH & 6.23 & 15.24 & 33.26 & 13.38 & 37.83 & 57.39 & 21.07 & 9.68 & 5.73\\
%SH & 6.87 & 15.93 & 34.40 & 14.32 & 38.63 & 58.33 & 21.98 & 10.03 & 6.16\\
%Weighted SH& 5.39 & 1.26 & 1.37 & 0.97 & 1.11 & 8.47 & 23.20 & 11.20 & 8.08\\
%\hline
%\end{tabular}
%\end{table}
%\doublespacing

\singlespacing
\begin{table}
\caption{Mean Squared Error (times $10^3$) of Simulated Prevalence Estimates on 1000 Networks with 1000 Nodes}
\label{tab:Homo_RE_comparison_all}
\begin{tabular}{l c |c c c |c c c c}
\hline \hline
Recruitment Effectiveness: & {1,1}& {1,1}& {1,1}& {1,1}& {1,1}&{1,1}& {0.9,0.6}& {0.9,0.6} \\ 
Homophily: & {1}& {1}& {1}& {1}& {1}& {2}& {1}& {2} \\
Differential Activity: & {1}& {2}& {2}& {2}& {2}& {2}& {2}& {2} \\
Sampling Fraction: & {0.2}& {0.2}& {0.5}& {0.7}& {0.2}& {0.2}& {0.2}& {0.2} \\
\hline 
Estimator & \multicolumn{8}{c}{}\\

Mean & 	0.69 & 13.65 &	6.47 & 2.72 & 13.65 & 18.40 & 13.95 & 36.00 \\
SS & 0.84 & 0.54 & 0.10 & 0.03 & 0.54 & 1.68 & 0.63 & 4.85 \\
VH & 0.88 & 0.62 & 1.52 & 3.33 & 0.62 & 1.50 & 0.62 & 3.30 \\
SH & 0.93 & 0.69 & 1.59 & 3.44 & 0.69 & 1.79 & 0.70 & 2.36 \\
Weighted SH & 0.90 & 0.54 & 0.13 & 0.14 & 0.54 & 1.62 & 0.63 & 1.71 \\
\hline
\hline
\end{tabular}
%\end{center}

\end{table}

\doublespacing

\begin{figure}[tbp] % float placement: (h)ere, page (t)op, page (b)ottom, other (p)age
  \centering
  % file name: C:/Users/Miles/Dropbox/Miles_Krista/Preparing RDS Re-Weighted Paper for Publication/pop-miss-specified-side_by_side_boxplots.pdf
  \includegraphics[bb=12 13 850 404,width=5.67in,height=2.65in,keepaspectratio]{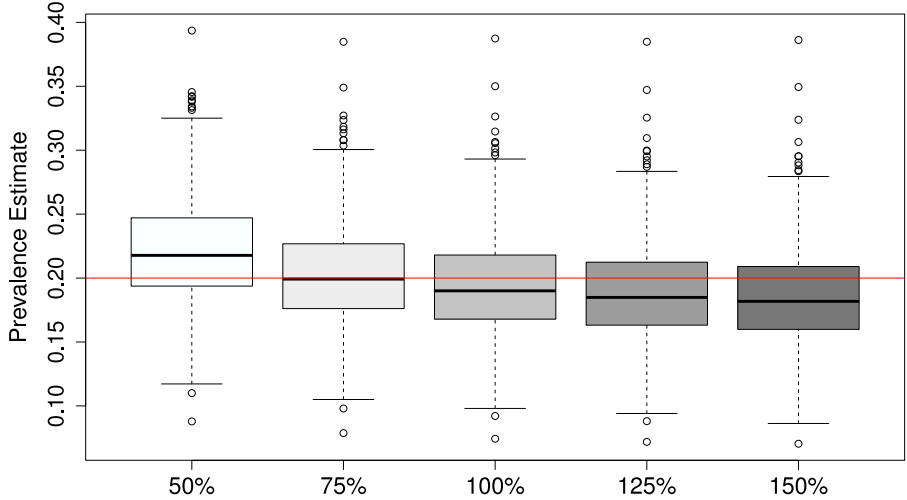}
  \caption{Prevalence estimates weighted SH estimators with homophily=2, recruitment effectiveness =(.9,.6), differential activity =2, with a sampling fraction of 20\% and varying the specification of the population size to be 50\%, 75\%, 100\%, 125\%, and 150\% of the original sample size of 1000. The horizontal red line represents the true prevalence.}
  \label{fig:side_by_side_boxplots_pop_misspec}
\end{figure}

In each of the previous simulations, it is assumed that the population size is known in the estimation for both the SS and the weighted SH. In the final set of simulations, we focus exclusively on the weighted SH estimator, and investigate the impact of misspecifying the population size. In these simulations, we again use 1000 networks where the population size is 1000, homophily and differential activity are both present.  We simulate respondent driven sampling on each network, using a sample size of 200, and induce differential recruitment effectiveness of (.9,.6).  We then calculate the weighted SH in each of these simulations varying the the specified population size from 50\% of the true population size to 150\% of the true population size by increments of 25\% which we display in Figure~\ref{fig:side_by_side_boxplots_pop_misspec}.  We found that the MSE for each of these estimates was less than the MSE for any of the other estimators under the same sampling and network conditions (see the last column in the in third panel of Table~\ref{tab:Homo_RE_comparison_all}), and that the MSE was lowest for the weighted SH with the correct sample size specified during the estimation process.

\section{Analysis of Mauritius Data}\label{app}

We applied this method to data sampled using RDS in 2011 among PWID in Mauritius.  Using six seeds, each participant was provided with up to three recruitment coupons resulting in a sample size of 500 and a maximum sample wave of 12.  Here we focus on estimating the prevalence of HIV and Hepatitis C, and the proportion of female PWID in Mauritius.  

\subsection{HIV Prevalence}

In the sample, 219 respondents were HIV positive, 279 were HIV negative, and 2 had missing values.  Here all missing values were coded as HIV negative. The average degree in the sample was 12.79, and the average degree among those who were HIV positive was 10.38, while the average degree among those who were HIV negative was 14.67.  HIV positive participants in the sample successfully recruited 1.05 on average, while HIV negative participants successfully recruited 0.93 on average (recruitment effectiveness ratio=1.14) \citep{GileDiagnost2015}.  As there seems to be both differential activity and differential recruitment effectiveness  and assuming that homophily is present in this sample, the weighted SH estimator developed in this work may be of use.  In the presence of differential recruitment effectiveness, homophily, and differential activity, we would expect the VH, SS, and SH to overestimate the proportion of the population that is a member of the group with smaller average degree.  In particular, since those who were HIV positive had lower average degree than those who were HIV negative, we would expect that the weighted SH estimator would provide a more accurate estimate of HIV prevalence, and that the other estimators would tend to overestimate HIV prevalence.  

In order to apply the weighted SH estimator, we must use an estimate of the total number of PWID in Mauritius.  We used the estimated population size of 9253 at the time of the study \citep{JohnstonMauritius}. Using this data set and the SH method we estimate:
$\widehat{C}{(AB)}=0.48$, $\widehat{C}{(BA)}=0.36$, $\widehat{D}{(A)}=4.46$,$\widehat{D}{(B)}=6.31$.  For the weighted SH method, we estimate $\widehat{C}{(AB)}=0.56$, $\widehat{C}{(BA)}=0.26$, $\widehat{D}{(A)}=4.31$, $\widehat{D}{(B)}=6.02$.    The estimated proportions of HIV positive PWID are 51.35\% (SH), 39.65\% (Weighted SH), 51.89\% (SS), 52.10\% (VH), and 43.80\% (naive mean).

\begin{figure}[tbp] % float placement: (h)ere, page (t)op, page (b)ottom, other (p)age
  \centering
  % file name: C:/Users/Miles/Dropbox/Miles_Krista/Preparing RDS Re-Weighted Paper for Publication/RDS_Mauritius_CIs.pdf
  \includegraphics[bb=4 2 476 476,width=2.74in,height=2.76in,keepaspectratio]{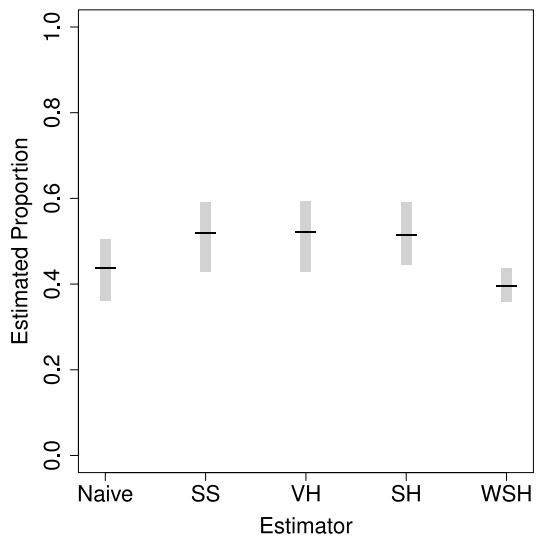}
  \caption{Estimated proportion of HIV positive PWID with 95\% CIs for five different estimators.}
  \label{fig:RDS_Mauritius_CIs}
\end{figure}

Figure \ref{fig:RDS_Mauritius_CIs} displays the estimated proportions of HIV positive PWID and a bootstrap 95\% confidence interval  (CI) %\citep{Salganik2006} 
(from 10,000 bootstrap samples) for all five estimators that we discuss here.  In the figure we can see that the weighted SH estimator has both the lowest estimate of the proportion who are HIV positive and the narrowest 95\% CI.  The VH has the highest estimate.  %These results are in line with what we would expect to see when there is differential recruitment effectiveness and differential activity with those who are HIV positive having lower average degree. 

\subsection{Hepatitis C Prevalence}

In the sample, 92.2\% respondents tested positive for the Hepatitis C virus. The average degree among those who are Hepatitis C positive was 12.90, while the average degree among those who were Hepatitis C negative was 11.46.  Hepatitis C positive participants in the sample successfully recruited 1.00 on average, while Hepatitis C negative participants successfully recruited 0.74 on average (recruitment effectiveness ratio=1.35).  Since those who tested positive for Hepatitis C had higher average degree than those who tested negative for Hepatitis C, we would expect that the weighted SH estimator would provide a more accurate estimate of Hepatitis C prevalence, and that the other estimators would tend to underestimate Hepatitis C prevalence.  

%In order to apply the weighted SH estimator, we must use an estimate of the total number of people who inject drugs in Mauritius.  As before when estimating HIV prevalence, 
As before, we used the estimated population size of 9253 and the SH method to estimate:
$\widehat{C}{(AB)}=0.08$, $\widehat{C}{(BA)}=0.93$, $\widehat{D}{(A)}=5.16$,$\widehat{D}{(B)}=4.84$.  For the weighted SH method, we estimate $\widehat{C}{(AB)}=.04$, $\widehat{C}{(BA)}=0.98$,$\widehat{D}{(A)}=5.39$,$\widehat{D}{(B)}=5.04$.  The estimated prevalence of PWID who are Hepatitis C positive are 91.63\% (SH), 95.79\% (Weighted SH), 91.75\% (SS), 91.74\% (VH), and 92.20\% (naive mean).

\begin{figure}[tbp] % float placement: (h)ere, page (t)op, page (b)ottom, other (p)age
  \centering
  % file name: C:/Users/Miles/Dropbox/Miles_Krista/Preparing RDS Re-Weighted Paper for Publication/RDS_Mauritius_CIs_HCV.pdf
  \includegraphics[bb=4 2 476 479,width=2.74in,height=2.76in,keepaspectratio]{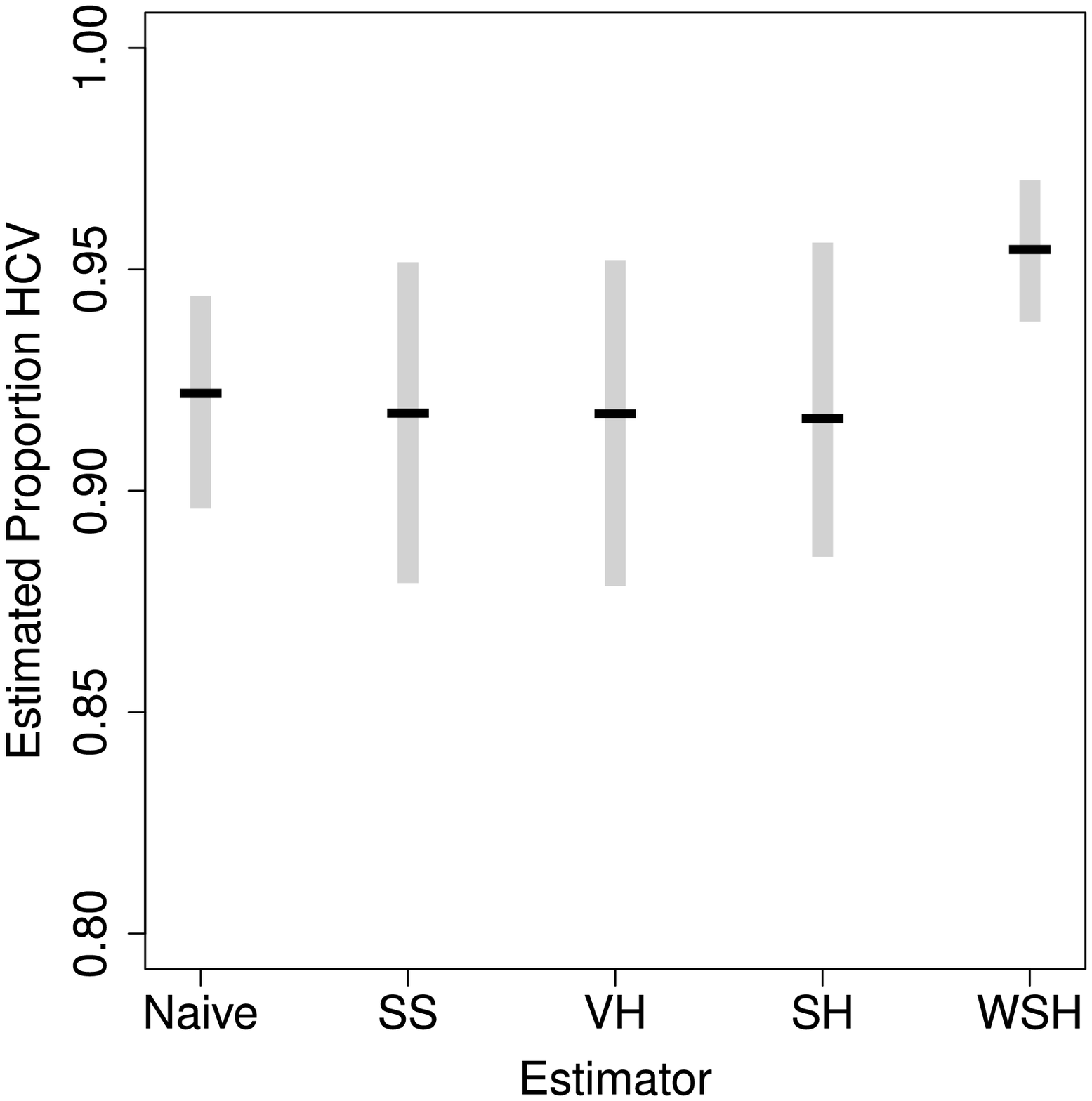}
  \caption{Estimated proportion of Hepatitis C positive PWID with 95\% CIs for five different estimators.}
  \label{fig:RDS_Mauritius_CIs_HCV}
\end{figure}

Figure \ref{fig:RDS_Mauritius_CIs_HCV} displays the estimated proportions of Hepatitis C positive PWID and a bootstrap 95\% CIs %\citep{Salganik2006} 
(from 10,000 bootstrap samples) for the five estimators.  In the figure we can see that the weighted SH estimator has both the
highest estimate of the proportion who are Hepatitis C positive and the narrowest 95\% CI. The
VH, SS, and SH all have comparable estimates. %These results are in line with what we would expect to see
when there is differential recruitment effectiveness and differential activity with those who
are Hepatitis C positive having higher average degree.

\subsection{Proportion Female}

In the sample, 6\% of respondents were female.  The average degree among those who were female was 13.77, while the average degree among those who were not female was 12.73.  Female participants in the sample successfully recruited 1 person on average, while non-female participants successfully recruited .98 people on average (recruitment effectiveness ratio=1.02).  

Contrasting the SH estimator and the weighted SH (as well as the other three estimators) when estimating the proportion of the population that is female, with the SH method we find: $\widehat{C}{(AB)}=0.83$, $\widehat{C}{(BA)}=0.05$, $\widehat{D}{(A)}=4.46$, $\widehat{D}{(B)}=5.18$.  For the weighted SH method, we estimate $\widehat{C}{(AB)}=0.83$, $\widehat{C}{(BA)}=0.03$, $\widehat{D}{(A)}=4.70$,$\widehat{D}{(B)}=5.42$.    The estimated prevalence of PWID who are female are 6.76\% (SH), 3.53\% (Weighted SH), 6.89\% (SS), 6.91\% (VH), and 6.00\% (naive mean). 

Since those who were female had higher estimated average degree (both with the SH method and the weighted SH method) than those who were not female, we would expect that the weighted SH estimator would provide a more accurate estimate of the proportion female, and that the other estimators would tend to overestimate the proportion female, which is what we concluded above.  

\begin{figure}[tbp] % float placement: (h)ere, page (t)op, page (b)ottom, other (p)age
  \centering
  % file name: C:/Users/Miles/Dropbox/Miles_Krista/Preparing RDS Re-Weighted Paper for Publication/RDS_Mauritius_CIs_Female.pdf
  \includegraphics[bb=4 2 476 479,width=2.74in,height=2.76in,keepaspectratio]{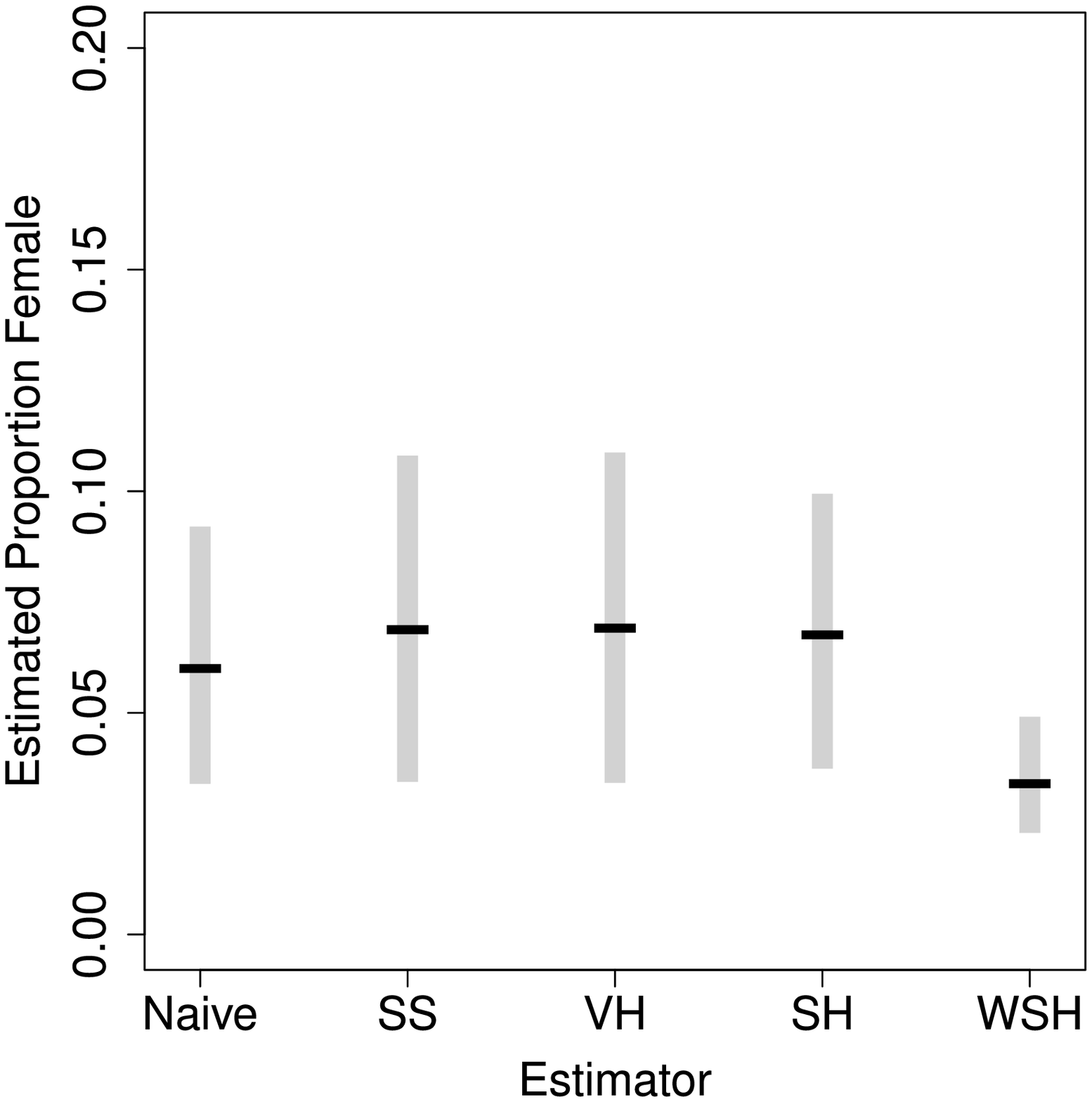}
  \caption{Estimated proportion of Hepatitis C positive PWID with 95\% CIs for five different estimators.}
  \label{fig:RDS_Mauritius_CIs_Female}
\end{figure}

Figure \ref{fig:RDS_Mauritius_CIs_Female} displays the estimated proportions of PWID who are female and a bootstrap 95\% CIs %\citep{Salganik2006} 
(from 10,000 bootstrap samples) for the five estimators. 

%\boxit{comment on statistical significance}
%\boxit{why do we expect to see this difference?  Is there differential recruitment effectiveness?  finite population effects?  How do we interpret this result?}

%\boxit{Below is stuff from NY Jazz data set}
%We also estimated the SS and weighted SH when specifying the underlying population to be 20,000, 30,000, and 40,000 since these two estimators depend on the underlying population size.  In this example, both the SS estimators were the same (80.18\% for all three different population sizes), as were the weighted SH estimator (84.31).  This indicates that the SS and weighted SH were not sensitive to the specified sample size.    

\section{Discussion}\label{dis}

In this paper, we have estimated features of the population of PWID in Mauritius.  To do so, we have introduced a new estimator which improves upon existing RDS prevalence estimation by accounting for the unequal edge sampling probabilities that result from the violation of the with-replacement sampling assumption.  Our new estimator, which we call the ``weighted SH'' weights edges inversely to an estimated edge sampling probability.  While this estimator follows the general form of the SH estimator \citep{Salganik2004}, it differs from the SH estimator in two important ways.  First, when edge characteristics are needed for the method of moments approach, rather than assume that all edges have an equal chance of being included in the RDS sample, we estimate the edge inclusion probabilities of each edge taking into account the without-replacement sampling inherent in RDS.  Estimating these edge inclusion probabilities is a difficult endeavor, and we have provided an approximation that allows % imperfect solution which tends to slightly overestimate these probabilities, but is a  %MQO:  I changed this sentence.
significant improvement over assuming all edge inclusion probabilities are equal.  Secondly, we also estimate the average degrees again accounting for the without-replacement sampling.  The weighted SH estimator will be particularly useful when the degree distributions and recruitment effectiveness vary by the outcome of interest, and homophily is present, conditions that are common in settings where RDS is typically used, including in each area of concern in our application in Mauritius (HIV, Hepatitis C status, and gender).

We have shown here that the weighted SH estimator that we propose has uniformly smaller MSE than the SH estimator in simulation, under several different conditions both of the network and of the sampling process.   Most notably, in the simulations which we think most closely represent real-world conditions (with differential recruitment effectiveness, homophily, and differential activity) the SH estimator has 1.38 times the MSE of the weighted SH, the VH has 1.93 the MSE of the weighted SH, and the SS has 2.83 times the MSE of the weighted SH.  Ideally, we would derive analytic results to compare estimators, however since the RDS is so complex, simulations provide the best option for comparing estimators to find the estimator that will perform the best in practice.  
 
While the weighted SH estimator that we propose here improves over the SH in many ways, the SH estimator does not require that the underlying population size is known.  Estimating the population size of hidden populations is a difficult task, though new statistical methodology can aid in this estimation \citep{handcock2014}. However, in simulation we found that even with large misspecification (50\% greater or 50\% less) of the population size, the weighted SH has lower MSE than all of the other estimators that we considered, including the SH and VH which do not assume that population size is known.  Like the SH estimator, the weighted SH estimator may be subject to biases introduced by biased seeds, preferential recruitment, or other sampling and network anomalies. The weighted SH estimator, like the original SH estimator tends to exhibit less bias than estimators that weight the nodal attributes (SS, VH, mean) in the presence of differential recruitment effectiveness coupled with homophily.  We also note that the SH estimator assumes that the underlying network is undirected, and our newly proposed estimator also makes this assumption. %However, as the weighted SH estimator is currently formulated, it assumes that there is no differential recruitment effectiveness when calculating the edge inclusion probabilities.  The weighted SH estimator could then be improved by incorporating estimates of differential recruitment effectiveness.   

We also propose an adaption of the commonly used bootstrap-based variance estimator \citep{Salganik2006}.  However other variance estimators have recently been proposed such as in \citet{baraff2016estimating}, which utilizes a tree bootstrap method, and shows promising results. Future research should investigate the performance of variance estimation methods when using different RDS estimation methods \citep{spiller2017evaluating}.  

RDS estimation is commonly applied to estimate HIV prevalence of traditionally higher-risk and hard to reach networked populations, such as men who have sex with men, sex workers, and PWID.  Here, we have applied this new estimator, as well as the most commonly used RDS estimators, to estimate HIV and Hepatitis C prevalence, and the proportion females among PWID in Mauritius.  This is an excellent case study, as the population is well-defined (Mauritius is an island country), and is one of the traditional hard to reach populations to which RDS prevalence estimations are applied.  In this study, males, those who were HIV positive, and those who were Hepatitis C negative had higher average degree than females, those who were HIV negative, and those who were Hepatitis C positive, respectively.  As a result of this differential activity, we would expect that the weighted SH would correct for a bias that previous estimators exhibit to overestimate the proportion of groups with lower average degree. Indeed, we have found that the most commonly used prevalence estimators over-estimate HIV prevalence and proportion female, and underestimate Hepatitis C prevalence in this population of PWID in Mauritius relative to the weighted SH.  The differences in prevalence estimates between the weighted SH and the other most commonly used RDS prevalence estimators could have substantial implications for disease prevention surveillance and policy.  

One of the main contributions of this work is to improve upon a widely adopted prevalence estimator for hidden and networked populations.  There are many different RDS prevalence estimators that are now in use.  No one estimator is superior in all settings.  %It is not apparent which estimator should be used in which setting.  Future work should find ways to identify the best estimator to be used in a given setting. 
We conclude that the new weighted SH estimator is best when there is differential activity, differential recruitment effectiveness, and homophily effects, which is what we would expect to see in a realistic network setting, and what we believe is present in the network of PWID in Mauritius.  %As such, we recommend the adoption of this new estimator.    

\section{Acknowledgments}

Research reported in this publication was supported by grants from NSF(SES-1230081), including support from the National Agricultural Statistics Service.  We are grateful to Ahmed Saumtally, Sewraz Corceal, Indrasen Mahadoo, and Farida Oodally, for allowing us to use these data.

\newpage
\bibliographystyle{plainnat}
\bibliography{RDS,RDS_try,RDS_try_again,bayesian_calibration,alcohol_references}

%%%%%%%%%%%%%%%%%%%%%%%%%%%%%%%%%%%%%%%%%%%%%%%%%%%%%%%%%%%%%%%%%%%%%%%%%%%%%%%%%%%%%%
\end{document}